\documentclass[12pt,reqno]{article}
\usepackage{amsmath}
\usepackage{amsthm}
\usepackage{graphicx,indentfirst}
\usepackage[psamsfonts]{amssymb}
\numberwithin{equation}{section}

\def \Fig#1#2#3 {
\begin{figure}
\centering
\epsfxsize=#2cm \epsfbox{#1.eps}
\caption{#3}
\label{#1}
\end{figure}
}

\newcount\figno
\def\fig#1#2#3{
\par\begingroup\parindent=0pt\leftskip=1cm\rightskip=1cm\parindent=0pt
\baselineskip=15pt
\global\advance\figno by 1
\epsfxsize=#3
\centerline{\epsfbox{#2}}
\vskip 12pt
{\bf \small Figure \the\figno:} {\small #1}\par
\endgroup\par
}
\def\figlabel#1{\xdef#1{\the\figno
\mbox{ }}}
\def\encadremath#1{\vbox{\hrule\hbox{\vrule\kern8pt\vbox{\kern8pt
\hbox{$\displaystyle #1$}\kern8pt}
\kern8pt\vrule}\hrule}}

\def\b{{\beta}}
\def\bb{\bar{\b}}
\def\c{{\gamma}}
\def\bc{\bar{\c}}
\def\pl{\partial}
\def\bpl{\bar{\pl}}
\def\bz{\bar{z}}

\def\brangle{\right\rangle}
\def\blangle{\left\langle}

\def\cig{{\text{cig}}}
\def\SL{{\text{SL}}}

\title{The FZZ-Duality Conjecture - A Proof}
\author{\\ Yasuaki Hikida$^{1}$ and  Volker Schomerus$^{2}$
\\[5mm]
$^1$ High Energy Accelerator Research Organization (KEK) \\ Tukuba, Ibaraki 305-0801, Japan
\\[5mm]
$^2$ DESY Theory Group, DESY Hamburg  \\
Notkestrasse 85, D-22603 Hamburg, Germany}

\date{May 2008}

\begin{document}
\begin{titlepage}      \maketitle      \thispagestyle{empty}

\vskip1cm
\begin{abstract}
We prove that the cigar conformal field theory is
dual to the Sine-Liouville model, as conjectured originally by
Fateev, Zamolodchikov and Zamolodchikov. Since both models possess
the same chiral algebra, our task is to show that correlations of
all tachyon vertex operators agree. We accomplish this goal through
an off-critical version of the geometric Langlands duality for
sl(2). More explicitly, we combine the well-known self-duality of
Liouville theory with an intriguing correspondence between the cigar
and Liouville field theory. The latter is derived through a path
integral treatment. After a very detailed discussion of genus
zero amplitudes, we extend the duality to arbitrary closed surfaces.
\end{abstract}

\vspace*{-16.9cm}\noindent {\tt {arXiv:0805.3931}}\\ {\tt {KEK-TH-1253}}\\ {\tt {DESY 08-062}}
\bigskip\vfill
\noindent
\phantom{wwwx}{\small e-mail: }{\small\tt
hikida@post.kek.jp, volker.schomerus@desy.de}
\end{titlepage}

\baselineskip=19pt


\setcounter{equation}{0}
\tableofcontents
\newpage

\section{Introduction}

Strong-weak coupling dualities in 2-dimensional quantum theory
possess a long history, with a remarkable range of applications.
In addition to providing indispensable tools for concrete
calculations, they have also taught us important lessons about
non-perturbative quantum field theory in general. Among these
dualities, those relating different target space geometries of
sigma models have received special attention, in particular from
string theorists. Many extensions of the famous $R \leftrightarrow
1/R$ duality for a compactified free bosonic field $X \sim X + 2
\pi R$ were found, see e.g.\ \cite{Giveon:1994fu} for a review of
developments in string theory and further references to original
research papers. Mirror symmetry of Calabi-Yau compactifications
has allowed to sum up contributions from world-sheet instantons, a
task that would seem virtually impossible without duality
symmetries.

Non-perturbative contributions to the $\alpha'$ dependence of
the world-sheet model are potentially important whenever the
target space of the string becomes strongly curved, or, more
generally, when some geometric length scale is of the order
of the string length $l_s \sim \sqrt{\alpha'}$. In recent
years, a new class of curved target spaces has come into focus.
Through the AdS/CFT correspondence, strongly
curved (asymptotically) Anti-de Sitter (AdS) backgrounds are
believed to encode interesting information about gauge field
theories. As simple examples show, the corresponding world-sheet
theories receive non-perturbative corrections. It would be of
obvious interest to capture those through perturbative expansions
in some dual field theory, whenever such a dual exists.

At the moment, very little is known about target space dualities
involving AdS or more general holographic backgrounds, with only
one exception. It is provided by the 2D Euclidean black hole, also
known as the (semi-infinite) cigar. The cigar conformal field
theory contains two fundamental fields, $\phi$ and $X$, which take
their values in the real line $\phi \in ] - \infty, \infty ]$ and
in the interval $X \in [0,2 \pi R]$, respectively. In these
coordinates, the non-trivial background metric $ds^2$ and the
dilaton $\Phi$ read
\cite{Elitzur:1991cb,Mandal:1991tz,Witten:1991yr}
\begin{equation}
ds^2 \ = \ k\, \frac{e^{-2\phi}}{1+e^{-2\phi}}\,
 ( d\phi^2 + k^{-1} d X^2) \ \ , \ \ \
e^{-2\Phi} \ = \ e^{-2\Phi_0} \, (1+ e^{-2\phi}) \ \ .
\label{cigarmetric}
\end{equation}
The compactification radius $R$ of the field $X$ is related to
the parameter $k$ in the metric by $k = R^2$. Throughout this
entire note we shall set $\alpha ' = 1$. The cigar geometry,
along with its Lorentzian counterpart and supersymmetric
extensions, possesses many interesting applications. For us, it
serves as the simplest holographic background in which duality
can be addressed (see e.g.\ \cite{KKK} and references therein).

Duality symmetries in quantum field theory should come with a
prescription that maps fields of one model onto those of its
dual. Let us recall that the tachyon vertex operators of the
cigar conformal field theory are associated with the asymptotic
data of wave functions at $\phi = - \infty$,%
\begin{equation}
\Psi^{\cig;\, j}_{(n,w)}(z,\bar z) \ = \
  e^{2b(j+1)\phi}\  e^{i\frac{n}{R}
  X + i R w \widetilde X} \ \ \mbox{ where } \ \
\widetilde X \ = \ \widetilde X(z,\bar z)
\ = \ - i \int^{(z,\bar z)}  * d X
\end{equation}
is dual to the field $X = X(z,\bar z)$. It is defined through
a line integral over the Hodge dual $*dX$ of the differential $dX$.
We have also introduced the parameter $b = 1/{\sqrt{k-2}}$
that will appear frequently throughout our entire presentation. The
parametrization of tachyon vertex operators in terms of the complex
radial momentum $j$ and the quantum numbers $n,w$ for the circle
direction follows the usual conventions.

According to a famous FZZ-conjecture of V. Fateev, Al.
Zamolodchikov and A.B. Zamolodchikov \cite{FZZ} (see also
\cite{KKK} for a review), a dual of the cigar conformal field
theory is given by the so-called Sine-Liouville model. This dual
theory also involves two fields $\phi$ and $X$ with values in
$\phi \in ]-\infty,\infty[$ and $X \in [0,2\pi R]$, as before. The
two coordinates parametrize a  cylinder with the trivial flat
metric and radius $R = \sqrt{k}$. The motion of strings towards
$\phi \sim  \infty$ is cut off by a tachyon potential of the form
\begin{equation} \label{SLV}
V(\phi,X) \ = \ 4 \pi \lambda \  e^{\frac{1}{b} \phi} \ \cos
(\sqrt{k} \tilde X)\ \ .
\end{equation}
The operator $V$ is marginal, provided the field $\phi$ has
background charge $Q_\phi = b$, the same as in the cigar model
above. Tachyon vertex operators in Sine-Liouville theory are again
parametrized by data in the asymptotic domain $\phi \rightarrow -
\infty$ where the interaction is exponentially suppressed, i.e.\
they take the same form as for the cigar
\begin{equation}
\Psi^{\SL;\, j}_{(n,w)} (z,\bar z) \ = \ e^{2b(j+1)\phi}
 \ e^{i \frac{n}{R} X
+  i { R w} \widetilde X}\ .
\end{equation}
The FZZ conjecture claims that one may identify vertex operators
in the cigar background and Sine-Liouville theory,
\begin{equation} \label{correq}
\left\langle\,  \prod_{\nu=1}^N \, \Psi^{\cig;\,
j_\nu}_{(n_\nu,w_\nu)} (z_\nu,\bar z_\nu) \
\right\rangle^\cig_{(k)} \ = \ {\cal N}  \left\langle\,
\prod_{\nu=1}^N \, \Psi^{\SL;\, j_\nu}_{(w_\nu,n_\nu)} (z_\nu,\bar
z_\nu) \ \right\rangle^\SL_{R} \ \ .
\end{equation}
The normalization ${\cal N}$ depends on the parameter $b$,
the winding number violation $S = \sum w_\nu$, and the number
$N$ of vertex operators. The
coupling constant $\lambda$ of Sine-Liouville theory relates to
the parameter $b$. An explicit formula will be spelled out below.

The FZZ-duality has several interesting features. To begin with,
it is a strong-weak coupling duality. In fact, the cigar conformal
field theory is weakly coupled for large $k$ or, equivalently, for
small $b$. In this regime, the tachyon potential \eqref{SLV}
increases rapidly towards larger values of $\phi$. Hence, the
model becomes strongly coupled. Another point worth stressing is
that the FZZ-duality relates a sigma model to another
2-dimensional field theory with constant metric and linear
dilaton, but non-trivial tachyon potential. The latter is a finite
sum of exponentials. In this respect, the relation between the
cigar and Sine-Liouville theory is very different from the
geometric target space dualities which are produced e.g.\ by the
Buscher rules \cite{Buscher1,Buscher2}. Some tests of the
FZZ-duality on the sphere were performed in \cite{FH}. A
supersymmetric version was established in \cite{Hori:2001ax}, but
their argument is passing through gauged linear sigma models and
hence rather indirect. Our aim here is to give a direct derivation
of the equality \eqref{correq}, first for the sphere and then on
an arbitrary surface.

Our proof involves two key ingredients. As a starting point,
we represent the cigar theory as a coset conformal field
theory, obtained by gauging a certain U(1) symmetry in the
$H^+_3$ WZNW model. The coset formulation then allows us to
invoke an intriguing correspondence between correlation
functions in the $H_3^+$ WZNW model and Liouville field
theory \cite{RT,HS}. Thereby, we shall be able to express
correlation functions of the cigar conformal field theory
through special correlators in a product of Liouville field
theory and a free boson. The additional bosonic field arises
from the gauge field of the coset construction. It
participates actively in the reduction from the cigar
to Liouville field theory. Up to this point, all steps
are performed in the path integral formulation of the cigar,
following closely our recent work \cite{HS}. Clearly, such
path integral manipulations are not sufficient to transfer
us from a weakly to a strongly coupled regime. This is where
the second central ingredient comes in. Let us recall that
Liouville field theory is self-dual, i.e.\ it looks exactly
the same at weak and strong coupling. Having expressed
correlators of the cigar through Liouville theory we can
capitalize on the self-duality of the latter to describe
the cigar in the strong coupling regime. The resulting
formulation of the cigar conformal field theory will not
look like Sine-Liouville theory at first, but the two
descriptions shall turn out to be related through simple
rotations and reflections in field space. We shall
describe these explicitly in the last part of our
derivation, following ideas from an unpublished note
of V. Fateev \cite{Fateev}.

The plan of our work is as follows. In the next section we
shall review and extend our previous path integral derivation
of the correspondence between the $H_3^+$ WZNW model and
Liouville field theory. The discussion will incorporate
sectors obtained by the action of spectral flow
\cite{Maldacena:2000hw}. Such an
extension was described by Ribault in \cite{Ribault} and
it is crucial for us in order to treat winding number
violating amplitudes of the cigar conformal field theory.
Our proof of eq.\ \eqref{correq} is then presented in section 3.
The fourth section contains a generalization of the FZZ-duality
and its proof for higher genus surfaces. In the conclusions we
finally present a list of open problems and possible
applications.

\section{The $H^+_3$-Liouville correspondence}
\def\cB{\mathcal{B}}

The main purpose of this section is to extend our path integral
derivation of the correspondence between the $H^+_3$ WZNW model
and Liouville field theory to sectors obtained through spectral
flow \cite{Maldacena:2000hw}. On the sphere, the corresponding
relation between correlation functions has been derived by
algebraic means in \cite{Ribault}. Generalizing the treatment of
\cite{HS}, we shall arrive at the same result. Our new derivation,
however, has two advantages. First of all, it also applies to the
case of maximal winding number violation that could not be treated
previously \cite{Ribault}. More importantly, our argument extends
to surfaces of higher genus. Those will be dealt with
in section 4.%
\smallskip

In deducing the main formula \eqref{R1} below, we shall sketch the
key ideas explained in \cite{HS}. As we are going through the
individual steps, we shall present them in a somewhat different
light, stressing the similarities with the standard derivation of
T-duality. Of course, the correspondence between the $H^+_3$ model
and Liouville field theory is not a T-duality, as e.g.\ both
theories possess different central charges. Nevertheless, through
this correspondence, Liouville theory manages to capture all
information about tachyon correlators in the $H^+_3$ model.  At
first sight, this might actually seem a bit surprising. While
tachyon vertex operators in the WZNW model carry a 3-component
target space momentum, tachyonic modes in Liouville field theory
possess momentum in one direction only. The resolution of this
apparent paradox is intriguing: Only the radial momentum of the
WZNW model is mapped to a momentum in Liouville theory. The two
remaining components of  target space momentum, on the other hand,
become parameters specifying the world-sheet insertion points for
degenerate fields in Liouville theory. Such additional insertions
are necessary precisely because the models on both sides of the
correspondence have different target space dimension (central
charge).

\subsection{Correlators in the $H^+_3$ WZNW model}

As is any derivation of T-dualities
(see, e.g. \cite{Buscher1,Buscher2}), our first step is to present
the $H^+_3$ WZNW model in a first order formulation. To this end,
we introduce two auxiliary fields $\beta$ and $\bar \beta$ of
weight $h=1$. These supplement three fields $\phi, \gamma$ and
$\bar \gamma$ of conformal weight $h=0$ that come with the target
space coordinates of the Euclidean $AdS_3$. The action of the
model reads
\begin{equation} \label{action}
S^{\text{WZNW}}_k[\phi,\c,\b] \ = \ \frac{1}{2\pi} \int d^2w \,
\left( \pl \phi \bpl \phi -  \b \bpl \c - \bb \pl \bc +
\frac{Q_\phi}{4} \, \sqrt{g} {\cal R}\, \phi - b^2 \b\bb
e^{2b\phi}\right) \ \ .
\end{equation}
In order for the interaction term to be marginal, the field $\phi$
must possess a background charge $Q_\phi = b = 1/\sqrt{k-2}$. The
usual WZNW model on $H^+_3$ may be recovered from eq.\
\eqref{action} by integration over $\beta$ and $\bar \beta$.

Our task is to compute correlation functions of tachyon vertex
operators. There exist several natural ways to parametrize the
space of tachyonic modes.  The choice we shall adopt is given by
\begin{align} \label{vertex}
 V_j ( \mu | z ) \equiv \, | \mu |^{ 2 j + 2}
 \, e^{ \mu \gamma(z) - \bar \mu \bar \gamma(\bar z)}\,  e^{2b (j +1 )
\phi(z,\bz)} \ \ .
\end{align}
These vertex operators are labeled directly by the three target
space momenta $\mu,\bar \mu$ and $j$. Our particular choice of
$\mu$-dependent prefactor will turn out to be very convenient
later on. The quantities we want to compute are the correlators
\begin{equation} \label{Sampl}
 \blangle \ \prod_{\nu=1}^{N} \, V_{j_\nu}(\mu_\nu|z_\nu) \
 v^S (\xi) \brangle^H \ = \ \int_{(S,\xi)} {\cal D} \phi {\cal D} \gamma
{\cal D}\beta \ e^{-S^{\text{WZNW}}_k[\phi,\gamma,\beta]}\,
\prod_{\nu = 1}^N V_{j_\nu}(\mu_\nu|z_\nu)  \ e^{S\phi(\xi)/b}\ .
\end{equation}
Here, the superscript $H$ reminds us to evaluate the correlation
function in the $H^+_3$ WZNW model. The operator $v^S(\xi)$ we
inserted at $z = \xi$ on the left hand side is obtained by acting
with $S$ units of spectral flow on the identity field. We let the
superscript $S$ run through positive integers. The generalization
to negative $S$ is quite obvious but dealing with both signs at
the same time would clutter our presentation below. The insertion
of $v^S(\xi)$ has two effects on the right hand side of eq.\
\eqref{Sampl}. To begin with, it leads to an insertion of the
vertex operator $\exp(S\phi/b)$. Moreover, $v^S(\xi)$ determines
the integration domain for the fields $\beta$ and $\bar \beta$ in
the path integral. To be more precise, the integration in eq.\
\eqref{Sampl} is meant to extend over all field configurations
such that $\beta$ and $\bar \beta$ both possess a zero of order
$S$ at $z=\xi$. In our analysis \cite{HS}, we had set the
parameter $S$ to $S=0$ and studied the usual path integral for
vacuum expectation values. With later applications in mind we now
extend this treatment.

Even though we may think of $v^S(\xi)$ as being defined through
the path integral representation we outlined in the previous
paragraph, it might be useful to pause for a moment and explain
the precise relation to the action of the spectral flow
automorphism $\rho^S$ on the affine sl(2) current algebra. In the
following discussion we shall set $\xi = 0$ and pass to an
operator formalism. Our freedom of choosing $S$ can then be
understood as the insertion of a state $| S \rangle$ that is
obtained from the vacuum $|0 \rangle$ through $S$ units of
spectral flow, i.e.%
$$
  \blangle \ \prod_{\nu=1}^{N} \, V_{j_\nu}(\mu_\nu|z_\nu) \ v^S(0) \
 \brangle^H
 \ = \ \langle0|\, \prod_{\nu=1}^{N} \, V_{j_\nu}(\mu_\nu|z_\nu)\,  |S\rangle \
 ,
$$ where $|S\rangle$ satisfies
\begin{equation} \label{SF0}
 \rho^S(J^a_n)|S \rangle \ = \ 0 \ \ \ \ \ \ \ \mbox{ for } \ \ \
  n \ \geq\  0\ ,\  a\  =\  3,\pm\ \ .
\end{equation}
For the reader's convenience we recall that the spectral flow
automorphism $\rho^S$ of the sl(2) current algebra is defined by
\begin{equation}
 \rho^S ( J^3_n ) \ = \ J^3_n - \frac{k}{2} S \delta_{n,0} \ \ \ ,
 \ \ \ \ \  \rho^S ( J_n^{\pm} )\  = \ J^{\pm}_{n \pm S} \ \ .
\end{equation}
We would like to rephrase the conditions \eqref{SF0} in terms of
the fields that appear in our action. To this end, we spell out
the usual free field realization of sl(2) currents,
\begin{eqnarray}
 J^-(z) & = &  \beta(z)  ~, \qquad
 J^3(z)\  =\  : \beta(z) \gamma(z) : - \ b^{-1} \partial \phi(z) ~,
 \\[2mm]
 J^+(z) & = & : \beta(z) \gamma ^2(z) : - \ 2 b^{-1} \gamma(z) \partial
 \phi(z)
+ k \partial \gamma(z) ~.
\end{eqnarray}
The construction of currents through $\beta,\gamma$ and $\phi$
implies that $|S\rangle = |S\rangle_{(\beta,\gamma)} \otimes
|S\rangle_{\phi}$ may be characterized through
\begin{align}
 \beta_{n - S } | S \rangle_{(\beta,\gamma)} \ = \ 0 \ \ \ , \ \ \
 \gamma_{n  + S} | S \rangle_{(\beta,\gamma)} \  =\  0\ \ \
  \mbox{ for } \ \ n \ \geq\  0\ \ .
 \label{vac}
\end{align}
Furthermore, the state $|S\rangle_\phi$ carries a non-vanishing
charge with respect to the zero mode of the field $\phi$, i.e.\
\begin{equation}  | S \rangle_\phi \ = \
   e^{\frac{S}{b} \phi(0)} | 0 \rangle_\phi \
\ .\end{equation} By now we easily recognise the description of
$v^S(\xi)$ we gave above. In fact, the state $|S \rangle $ creates
a zero of order $S$ in the field $\beta(w)= \sum \beta_n w^{-n-1}$
at $w =0$. The effect of $|S\rangle$ on the field $\phi$ is
captured by the insertion of the vertex operator $\exp(S\phi/b)$.
Obviously, the insertion point of $v^S$ can be moved from $w =
0$ to any point on the sphere (or surface). The amplitudes
\eqref{Sampl} we considered here are $(N+1)$-point functions
containing $N$ unflowed states in addition to the state at
$z=\xi$. The latter is obtained from the identity by $S$ units of
spectral flow as mentioned before. More general correlators for which the $S$ units
are distributed among all $N+1$ fields are rather easy to find, as
discussed in \cite{Ribault}. The relation between the cigar and
Liouville field theory would be derived from its simplest form
when all the spectral flow is assembled in one insertion point.

\subsection{The correspondence with Liouville theory}

Since vertex operators do not depend on $\beta$ and $\bar\beta$,
these fields can easily be integrated out. The resulting action is
that of the WZNW model for the usual coordinate fields
$\gamma,\bar \gamma$ and $\phi$. A ``dual'' description of the
WZNW model emerges when we integrate out $\gamma$ and $\bar
\gamma$ instead of $\beta$ and $\bar \beta$. As in the case of
T-dualities, the integration over $\gamma$ and $\bar \gamma$ gives
constraints on $\beta$ and $\bar \beta$. Solutions to these
constraints are inserted back into the action. Thereby, we arrive
at the dual formulation we are after. Let us now see how all this
works out for our $H^+_3$ WZNW model.

As explained in much detail in \cite{HS}, integration over
$\gamma$ and $\bar \gamma$ gives the constraints
\begin{equation} \label{plb}
\bpl \b(w) \ = \ 2\pi \sum_{\nu=1}^N \mu_\nu \delta^2(w-z_\nu) \ \
\ \ , \ \ \ \ \pl \bar\b(\bar w) \ = \
  -  2\pi \sum_{\nu=1}^N \bar \mu_\nu \delta^2( w- z_\nu)\ .
\end{equation}
If it were not for the insertion of vertex operators, these would
simply require the one-differentials $\beta$ and $\bar \beta$ to
be (anti-)holomorphic. The vertex operators act like sources and
force $\beta$ and $\bar \beta$ to possess first order poles with
residues $\mu_\nu$ and $- \bar \mu_\nu$ at the insertion points,
respectively. On the complex sphere, a meromorphic
one-differential with these properties is uniquely determined to
be of the form
\begin{equation} \label{beta}
\b(w) \ = \  \sum_{\nu=1}^N \frac{ \mu_\nu}{w-z_\nu}\ \ ,
\end{equation}
and similarly for $\bar \beta$. Due to the insertion of
$v^S(\xi)$, $\beta$ and $\bar \beta$ possess a zero of order $S$
at $w=\xi$. Therefore, the parameters $\mu_\nu$ must obey the
following $S+1$ equations
\begin{equation}
\sum_{\nu=1}^N \frac{\mu_\nu}{(\xi - z_\nu )^n} \ = \ 0 \ \ \ \ \
\mbox{ for } \ \ \ \ n = 0,1,2,\dots,S\ .
\label{constraints}
\end{equation}
The first equation with $n=0$ ensures that $\b(w)$ has no pole at
$w=\infty$. The uniqueness of the solution to eqs.\ \eqref{plb} is
a new feature of our analysis, distinguishing it from the case of
usual T-dualities. In the standard cases, solutions to the
constraints are parametrized by a dual field. Thereby, field
theories related by an ordinary T-duality possess the same number
of fields. Because the solutions to our constraint equations
\eqref{plb} are unique, the resulting ``dual'' of the $H^+_3$ WZNW
model will have two fields less than the theory we started with.

The next step is to insert the solutions to the constraints back
into the action. This leaves us with a theory of a single field
$\phi$ and a Liouville-like interaction term. A second glance at
the resulting action, however, reveals an unpleasant feature: In
the place of Liouville's cosmological constant we find a rather
complicated function $|\beta(w)|^2$ depending on the insertion
points $z_\nu$ and momenta $\mu_\nu,\bar \mu_\nu$. We can try to
resolve this issue by absorbing the unwanted function
$|\beta(w)|^2$ into a shift of the zero mode of $\phi$. Since we
are going to shift $\phi$ by the logarithm of $\beta(w)\bar
\beta(\bar w)$, it is advantageous to bring $\beta(w)$ into a
product form first. Let us recall that for any one-differential,
the number of poles exceeds the number of zeroes by two. Hence,
$\beta(w)$ must have $N-2$ zeroes. Since we inserted the operator
$v^S(\xi)$, $S$ of these zeroes must come together at $w=\xi$. We
will denote the remaining $N-2-S$ locations of zeroes on the
sphere by $w=y_i$.%
\footnote{In this way we have shown that the total spectral flow
number must be less than $N-2$, i.e., $S \leq N-2$. The same
conclusion was reached from a group theoretic argument in appendix
D of \cite{MO3}.} Furthermore, a differential is uniquely
characterized by the position of its zeroes and poles up to an
overall factor $u$. Consequently, we can rewrite $\beta(w)$ in the
form
\begin{equation} \label{bspb}
\beta(w) \ = \ \ u \, \frac{(w-\xi)^S \prod_{i=1}^{N-2-S}
(w-y_i)}{\prod_{\nu=1}^N (w-z_\nu)} \ =: \ u \cB(w)  \ \ .
\end{equation}
Thereby, we have now replaced the $N$ parameters $\mu_\nu$ subject
to constraints \eqref{constraints} through $N-2-S$ coordinates
$y_i$ and a global factor $u$.
Now we are ready to introduce the
new bosonic field $\varphi$ through
\begin{equation} \label{phired}
\varphi \ := \ \phi + \frac{1}{2b} \left( S \ln|w-\xi|^2 +
 \sum_{i=1}^{N-2-S}
\ln |w-y_i|^2 - \sum_{\nu=1}^N \ln |w-z_\nu|^2
 - \ln |u\rho (w) |^2 \right) \ \ ,
\end{equation}
where the term in brackets is $\ln |u\mathcal{B}|^2$. Here we have
included a non-trivial Weyl factor $\rho(z)$. With this factor,
the world-sheet metric and its curvature are given as $ds^2 =
|\rho (z)|^2 dz d\bar z$ and $\sqrt{g} {\cal R} = - 4 \partial
\bar \partial \ln |\rho|$. Throughout most of the present note we
set $\rho (z) = 1$. But several details of the duality between the
sigma model and Liouville theory require a more careful treatment.
This applies in particular to the derivation of the shift in the
background charge and to a proper regularization $\lim_{w \to z}
|w - z|^2 = - \ln |\rho (z)|^2$ of the divergent expression
$\lim_{w \to z} |w - z|^2$, see \cite{HS} for details.

Through our redefinition \eqref{phired} of the field $\phi$ we
remove the factors of $\mu$ in the definition \eqref{vertex} of
the vertex operators,
\begin{equation} \label{VOred}
|\mu_\nu|^{2(j_\nu+1)} \, e^{2b(j_\nu+1)\phi(z_\nu)} \ = \
   e^{2b(j_\nu+1)\varphi(z_\nu)}
\end{equation}
and thereby all explicit $\mu$ dependence. It remains to rewrite
the kinetic term through the new field $\varphi$. Since $\partial
\bar \partial \phi $ and $\partial \bar \partial \varphi$ differ
by a bunch of $\delta$-functions which are localized at the points
$z_\nu$, $y_i$ and $\xi$, we obtain extra insertions of vertex operators
$\exp (\pm \frac{1}{b} \varphi)$ at the zeroes and poles of
${\mathcal B}$. The vertex operators inserted at $z_\nu$ combine
with the original tachyon vertex operators while those at $y_i$
are new. Similarly, there is an extra insertion of the operator
$\exp(-S\varphi/b)$ at the point $w=\xi$. It combines with the
vertex operator
\begin{align}
 e^{ S \phi(\xi)/b}\ =\
 |u \tilde {\cal B} (\xi) |^{-S/b^2}\ e^{S \varphi(\xi) /b}
 \quad ~,
 \qquad \quad \tilde {\cal B} (\xi) \equiv
\frac{\prod_{i=1}^{N-2-S} (\xi - y_i)}
 {\prod_{\nu = 1}^N (\xi - z_\nu)}
\end{align}
into some simple numerical factor $|u \tilde {\cal B} (\xi)
|^{-S/b^2}$. The latter is canceled by the numerical,
$\xi$-dependent term in $\phi \partial\bar \partial \phi$ so that
the end results assume the form
\begin{align} \nonumber
& \left\langle \prod_{\nu = 1 }^N
 V_{j_\nu} (\mu_\nu | z_\nu )
 v^S(\xi)\right\rangle =
  \prod_{n=0}^S \delta^2 \left(\sum_{\nu=1}^N \frac{\mu_\nu}{(\xi-z_\nu)^n} \right)
   \frac{| \Theta_N| ^2}{ |u|^{\frac{S}{b^2}-2}}
  \left\langle \prod_{\nu = 1}^N
 V_{\alpha_\nu} (z_\nu )\!\!\! \prod_{i=1}^{N-2-S}
 \!\!\!\! V_{-\frac{1}{2b}} (y_i)
 \right\rangle ^L  \\[3mm]
&\mbox{with }\quad
 \label{R1}
 \Theta_N  =  \Theta_N (u,y_j,z_\nu )  =   \prod_{\mu <
\nu}^N (z_{\mu \nu})^{\frac{1}{2b^2}} \prod_{i < j}^{N-2-S} (y_{ij})^{\frac{1}{2b^2}} \prod_{\nu =1 }^N\prod_{i =1}^{N-2-S}
 (z_{\nu} - y_i)^{- \frac{1}{2b^2}} ~.
\end{align}
Note that all dependence on the insertion point $z=\xi$ has
dropped from all terms but those implementing the constraints
\eqref{constraints}. The right hand side of eq.~\eqref{R1} is
evaluated in the Liouville theory. The vertex operators are
$V_{\alpha} = \exp (2\alpha\varphi )$ with $\alpha_\nu =
b(j_\nu+1) + 1/2b$, and the number of degenerated fields
$V_{-1/2b}$ is given by $N-2-S$.

All the above can be generalized to world-surfaces of higher genus
$g \geq 1$, as shown in section 4, following our analysis in \cite{HS}. The main point to
note concerns the number of additional insertions: On a surface of
genus $g$ a one-differential with $N$ poles possesses $N+2g-2$ zeroes. Once more,
$S$ of them should come together at the point at which we insert
the spectral flow of the identity field. The remaining $N+2g-2-S$
zeros give rise to the insertion of degenerate fields. With the
generalization to higher genus surfaces being well under control,
the relation between the $H^+_3$ WZNW model and Liouville field
theory becomes a perturbative correspondence that works order by
order in the string loop expansion.

\section{The Cigar--Sine-Liouville duality}
\def\cig{{\text{cig}}}

We are now ready to derive the duality between the cigar conformal
field theory and the Sine-Liouville model. Our argument proceeds
in several steps. First we use the correspondence between the
$H^+_3$ model and Liouville field theory to establish a similar
correspondence between the cigar and a new model that involves a
Liouville field $\varphi$ along with a single free boson $\chi$.
As before, the Liouville correlation functions contain $N-2-S$
additional degenerate field insertions. In this form, our
correspondence does not yet resemble the duality we were seeking
for. To begin with, the Liouville field theory with interaction
$\exp 2b\varphi$ remains weakly coupled for small $b$, i.e.\
whenever the $H^+_3$ WZNW model is weakly coupled. Furthermore,
the correspondence relates correlation functions with a different
number of field insertions. Finally, the background charges of the
Liouville field $\varphi$ and the boson $\chi$ are found to differ
from those in the Sine-Liouville model. We shall address each of
these three shortfalls in a separate subsection.

\subsection{A correspondence between the cigar and Liouville theory}

Our first aim is to determine correlation functions of tachyon
vertex operators in the cigar conformal field theory. As before,
we parametrize the cigar geometry through the radius $R= \sqrt k$
of the circle at $\phi = - \infty$. The value of $R$ determines the
central charge through
$$c_\cig \, = \, 2 \ \frac{k+1}{k-2}\ \ . $$
Let us recall that the cigar conformal field theory may be
obtained from the $H_3^+$ WZNW model by a process of gauging.
Thereby, the cigar model gets embedded into the combination of a
$H^+_3$ WZNW model at level $k$ and a free bosonic field theory,
$$
S[\phi,\gamma,\beta;X;b,c] \ = \
  S^{\text{WZNW}}_k[\phi,\gamma,\beta] \ + \ \frac{1}{2\pi}
 \int d^2w \ \partial X \bar \partial X
 \ + \  \frac{1}{2 \pi} \int d^2 w
 ( b \bar \partial c + \bar b \partial \bar c ) \ \
$$
along with $(b,c)$-ghosts arising from the gauge fixing procedure
(for more detail, see appendix \ref{cigarg}). The first term represents the
WZNW model, written once more in a first order formulation
\eqref{action}. The free boson $X$ is compactified to a circle of
radius $R = \sqrt{k}$, i.e.\ the compactification radius of $X$ is
the same as for the cigar at $\phi = -\infty$. It has vanishing
background charge $Q_X = 0$.

Next we turn our attention to the vertex operators.  Our
conventions for vertex operators of the WZNW model can be found in
\eqref{vertex}. In order to spell out the relation with vertex
operators of the cigar, we need to pass to the so-called $m$-basis
\begin{align}
 \Phi^j_{m,\bar m}(z)\  = \ N^j_{m,\bar m}
  \int \frac{d^2 \mu}{|\mu|^2}\,  \mu^{m} \bar \mu^{\bar m}
\, V_j (\mu | z) ~, \qquad
 N^j_{m,\bar m}\, =\,  \frac{\Gamma(-j-m)}{\Gamma (j+1+\bar m)} ~.
\end{align}
We have to combine these with vertex operators of the free boson
$X$. For the latter we use the following notation
\begin{equation}
V^X_{m,\bar m}(z,\bar z) \ = \ e^{i\frac{2}{\sqrt{k}}(m X_L
  - \bar m  X_R )}\ \ .
\end{equation}
Here, we have also introduced the fields $X_L = X_L(z,\bar z)$ and
$X_R = X_R(z,\bar z)$ through $X = X_L+ X_R$ and $\widetilde X =
X_L- X_R$. Note that our sign convention for $\bar m$ deviates
from the standard one. Vertex operators for the cigar are
constructed according to the simple rule
\begin{align}
 \label{parafield}
 \Psi^j_{m,\bar m}(z,\bar z) \ = \ V^X_{m,\bar m}(z,\bar z) \,
   \Phi^j_{m,\bar m}(z,\bar z)\ \ .
\end{align}
The two parameters $m$ and $\bar m$ denote the left and right U(1)
charges. They are related to the asymptotic momentum and winding
numbers $n$ and $w$ (see introduction) through $m = (kw+n)/2$ and
$\bar m = (kw-n)/2$.

Combining the results of the previous paragraphs, we are led to
the following basic representation of our correlation functions,
\begin{align} \label{corr}
 \left\langle \prod_{\nu = 1}^N
 \Psi^{j_\nu}_{m_\nu ,\bar m_\nu } (z_\nu ) \right\rangle
 ^{\text{cig}}
 & = \  \prod_{\nu =1}^N \left[ N^{j_\nu}_{m_\nu ,\bar m_\nu }
  \int \frac{d^2 \mu_\nu }{|\mu_\nu |^2}
  \mu_\nu^{m_\nu} \bar {\mu}_\nu^{\bar m_\nu} \right]
  \times \\[2mm] & \ \ \ \times
 \left\langle \  V^X_{-\frac{kS}{2},-\frac{kS}{2}}  (\xi) v^S (\xi)
 \prod_{\nu = 1 }^N
 \, V^X_{m_\nu,\bar m_\nu}(z_\nu) \
  V_{j_\nu} (\mu_\nu | z_\nu )
\ \right \rangle^{H\times F}\ .  \nonumber
\end{align}
The correlator on the right hand side is to be evaluated in a
product of the $H^+_3$ WZNW model with a single free boson. The
parameter $S$ that determines the insertion at $z=\xi$ is related
to the integers $m_\nu$ and $\bar m_\nu$ through $\sum_\nu m_\nu =
\sum_\nu \bar m_\nu = \frac{kS}{2}$. For the cigar conformal field
theory, the operator at $z = \xi$ is just a representation of
identity field. Hence, the right hand side of eq.\ \eqref{corr}
should not depend on the insertion point $z=\xi$, a property we
shall confirm explicitly below.

Our first step now is to insert the results from section 2 for the
correlation functions in the $H^+_3$ WZNW model. Thereby, we bring
in the correlators of Liouville theory,  multiplied by the rather
complicated prefactor $\Theta_N$ (see eq.\ \eqref{R1}). But there
remains some explicit $\mu$-dependence in the integrand along with
the integration over $\mu_\nu$. According to our general strategy,
we would like to rewrite the expressions entirely in terms of the
new variables $u$ and $y_i$. This works out very nicely if we
redefine the bosonic field $X$ in a way that is reminiscent of
what we did in eq.\ \eqref{phired} for the field $\phi$,
\begin{equation} \label{Xred}
\chi_L \ := \ X_L -  i \frac{\sqrt{k}}{2}\,
   \left(S \ln ( w-\xi ) +  \sum_{i=1}^{N-2-S} \ln(w-y_i)
   - \sum_{\nu = 1}^N \ln (w - z_\nu)   - \ln u\rho (w) \right)\ \ .
   \end{equation}
The field $\chi_R$ is defined by trading $X_L$ for $X_R$ and
taking the complex conjugate of the second term. In this way we
ensure that the local field $\chi(z,\bar z) = \chi_L + \chi_R$
remains real. Let us also note that its dual field $\tilde \chi=
\chi_L - \chi_R$ acquires a non-zero background charge $Q_{\tilde
\chi} = - i \sqrt k$. Therefore, the free bosonic field $\chi$ has
central charge $c_{\chi} = 1 - 6k$. Using the same reasoning as in
\cite{HS} we obtain
\begin{equation}
(\mu_\nu)^{m} (\bar \mu_\nu )^{\bar m}
 \, V^X_{m,\bar m}(z_\nu) \ = \
 V^\chi_{m,\bar m} (z_\nu) \ .
 \end{equation}
Thereby, we now got rid of all the explicit $\mu$-dependence in
the integrand. Our redefinition of the bosonic field also leads to
additional insertions of bosonic vertex operators into the
correlation functions, much in the same way as for the field
$\varphi$.

But there is one additional important consequence of the shift
\eqref{Xred}. It also produces a numerical factor similar to
$\Theta_N$, only with the exponent $1/b^2$ being replaced by $-k$.
Remarkably, the product of $\Theta_N$ with this new factor
combines exactly into the Jacobian for the transformation from
$\mu_\nu$ to $u,y_i$.
The latter is computed in Appendix \ref{jacobian}
and it reads
\begin{equation}
 \prod_{\nu=1}^N \frac{d^2 \mu_\nu}{|\mu_\nu|^2}
 \prod_{n=0}^S \delta ^2 \left(\sum_\nu \frac{\mu_\nu}{(\xi - z_\nu)^n} \right)
 \, = \,
   \frac{\prod_{\mu < \nu}^{N} |z_{\mu\nu}|^2
    \prod_{i<j}^{N-2-S} |y_{ij}|^2}{\prod_{\nu=1}^N
    \prod_{i=1}^{N-2-S} |z_\nu - y_i|^2}
\frac{d^2 u}{|u|^{4 + 2S}}
 \prod_{i=1}^{N-2-S} d^2 y_i~.
 \label{jacobxi}
\end{equation}
In applying this substitution rule, one has to be a bit careful.
Note that the parameters $\mu_\nu$ are not effected if we permute
the insertion points $y_i$. This means that our transformations
map the space of $\mu_\nu$ to a $(N-2-S)!$-fold cover of the $y_i$
coordinate hyper-plane. Putting all this together we finally
obtain
\begin{align}
  \left \langle \prod_{\nu = 1}^N \Psi^{j_\nu}_{m_\nu , \bar
m_\nu}
 \right\rangle^\cig
  & = \
  \int \frac{\prod_{i=1}^{N - 2 - S} d ^2 y_i}{(N-2-S)!}
  \prod_{\nu = 1}^N  N^{j_\nu}_{m_\nu , \bar m_\nu}
    \times \nonumber \\[2mm]
&  \times \left \langle\  \prod_{\nu=1}^N V_{\alpha_\nu} (z_\nu  )
 \, V^\chi_{m_\nu- \frac{k}{2},\bar m_\nu - \frac{k}{2}}(z_\nu)\
  \prod_{i =1}^{N-2 - S} V_{- \frac{1}{2b}} (y_i  )
 \, V^\chi_{\frac{k}{2}, \frac{k}{2}}(y_i)\, \right  \rangle \ \ .
\label{cigL}
\end{align}
The correlation function on the right hand side is evaluated in
the theory
\begin{equation}
\label{act0}  S(\varphi,\chi) \ = \
\frac{1}{2\pi} \int d^2w \left(
\partial \varphi \bar \partial \varphi + \partial \chi \bar
\partial \chi +
  \frac{\sqrt g}{4} {\cal R} (Q_\varphi \varphi +
  Q_{\tilde \chi} \tilde \chi)
 + b^2 e^{2b\varphi}
    \right) \ \ ,
\end{equation}
with background charges $Q_\varphi = b + 1/b$ and $Q_{\tilde \chi}
= - i\sqrt k$. We have thereby achieved our first goal, namely to
express $N$-point correlation functions in the cigar conformal
field theory through correlators of $2N-2-S$ fields in a product
of the Liouville model with a single free bosonic field. This is
as far as the $H_3^+$-Liouville correspondence can take us.

\subsection{Derivation of the duality with Sine-Liouville theory}

The correspondence we derived in the previous subsection falls
short of being a true duality for a number of reasons. To begin
with, it relates correlators in the weakly coupled cigar conformal
field theory to correlation functions in weakly coupled Liouville
theory. Here is where the famous self-duality of Liouville theory
comes to our rescue. Through the correspondence \eqref{cigL} it
provides us with a non-perturbative completion of the cigar
conformal field theory, i.e.\ a well defined prescription to
calculate cigar correlators for small values of the level $k$
(large $b$). The next unpleasant feature of our correspondence is
that it relates correlators with a different number of vertex
operators. Since the central charge of the theory \eqref{act0} is
the same as for the cigar conformal field theory, one may expect
to do better. Indeed, our correspondence shall be rewritten as a
duality between $N$-point correlation functions in the second
subsection. At that point we could have decided to stop if we were
not fully determined to recover Sine-Liouville theory. We shall
succeed in the last part of this subsection through a rotation in
field space and an appropriate field identification.

\subsubsection{A weak-strong coupling correspondence}

As we have just stated, the correspondence we obtained in the
previous subsection does not seem very useful yet: Both the cigar
and Liouville field theory get weakly coupled for small values of
the parameter $b$ (or $k \sim 2$). Put differently, the Liouville
interaction term $\exp (2b\varphi)$ becomes large in the regime of
small curvature radius $\sqrt{k}$ that we were hoping to describe.
Our path integral manipulations could not have given us anything
more. They capture the perturbative aspects of the two models and
hence relate the weakly coupled regimes of the $H^+_3$ (or the
cigar) and Liouville field theory. What makes such a
correspondence so useful is the fact that quantum Liouville theory
looks the same at strong and weak coupling
\cite{Zamolodchikov:1995aa}. There is no way to derive this
self-duality of Liouville field theory within the path integral
treatment. But since the Liouville theory is solved (see e.g.\
\cite{Schomerus:2005aq} for a review and references), its
self-duality under the reflection $b \rightarrow b^{-1}$ is fully
established \cite{Ponsot:1999uf,Teschner:2003en}.

If we are ready to accept this additional input from quantum
Liouville theory, we can now compute our correlation functions in
the dual theory
\begin{equation} \label{act1}  S(\varphi,\chi) \ = \
\frac{1}{2\pi} \int d^2w \left( \partial \varphi \bar \partial
\varphi + \partial \chi \bar
\partial \chi +
  \frac{\sqrt g}{4} {\cal R} (Q_\varphi \varphi +
  Q_{\tilde \chi} \tilde \chi)  + \tilde \mu
  e^{\frac{2}{b}\varphi}
  \right) \ \ . 
\end{equation}
The background charge $Q_\varphi= b + 1/b$ of the Liouville field
remains the same as before. However, the inversion of the
parameter $b$ should be accompanied by an appropriate adjustment
of the bulk cosmological constant. In our case, the correct choice
is (see e.g.\ \cite{Zamolodchikov:1995aa,Teschner:2001rv})
\begin{equation} \label{mutilde}
\tilde \mu \ = \ \gamma^{-1}(b^{-2})
 \left( b^2 \gamma(b^2)\right)^{b^{-2}}
 ,
\end{equation}
where $\gamma(x) = \Gamma(x)/\Gamma(1-x)$, as usual. Let us stress
that correlation functions of the cigar conformal field theory are
still calculated through equation \eqref{cigL}. There is no need
to change the parameters of vertex operators, in spite of the fact
that they happen to be functions of $b$. After the inversion of
$b$, the interaction term $\exp 2\varphi/b$ becomes weakly coupled
when we enter the strong coupling regime of the cigar conformal
field theory.

\subsubsection{Removing degenerate field insertions}

When we were discussing the correspondence between the $H^+_3$
WZNW model and Liouville theory we argued that degenerate field
insertions were required in order to encode all information about
the target space momenta on $H^+_3$. The situation has changed
now. By gauging one of the directions of the $H^+_3$ model we
descended to a 2-dimensional target space. An $N$-point function
on the cigar involves only $2N$ target space momenta and hence the
$N-2$ insertion points in Liouville theory are certainly more than
is needed to simply store information on target space momenta.
Since we kept the bosonic field $X$ in our theory rather than
integrating it out, it should even be possible to do without any
additional field insertions.

This is indeed the case, due to the following observation
\cite{Fateev,Giribet}: The integrated insertions at the points
$y_i$ appear as if they had come from expanding an additional
interaction term in the action with the field
\begin{equation} \label{int}
V_{- \frac{1}{2b}}(y) \, V^\chi_{\frac{k}{2},\frac{k}{2}}(y) \ = \
e^{- \frac{1}{b}\, \varphi(y,\bar y) + i \sqrt k \, \tilde \chi(y,
\bar y)}\ ,
\end{equation}
where $\tilde \chi(y, \bar y) = \chi_L - \chi_R$ as before. In
fact, the total charge of exponentials of the field $\chi$ in our
correlator is
$$ \frac{1}{\sqrt k}
\sum_{\nu=1}^N m_\nu - N \frac{\sqrt{k}}{2} +
\frac{\sqrt{k}}{2}(N-2 -S) \ = \ - iQ_\chi\ \ .
$$
Hence, if we expand the exponential of integrated vertex operators
of the form \eqref{int}, only a single term contributes, namely
the one with $N-2-S$ insertions of the interaction. Here it is
essential that $\chi$ is a compact free bosonic field. Thereby, we
have shown that correlation functions in the cigar conformal field
theory
\begin{equation} \label{cigL1}
 \left \langle \prod_{\nu = 1}^N \Psi^{j_\nu}_{m_\nu , \bar
m_\nu}
 \right  \rangle^\cig
  \ = \ \pi^{N-2-S}
  \prod_{\nu = 1}^N  N^{j_\nu}_{m_\nu , \bar m_\nu}
 \,  \left \langle\  \prod_{\nu=1}^N V_{\alpha_\nu} (z_\nu  )
 \, V^\chi_{m_\nu- \frac{k}{2},\bar m_\nu - \frac{k}{2}}(z_\nu)\
 \, \right  \rangle
\end{equation}
may be computed by evaluating the correlator on the right hand
side in the theory \begin{equation} \label{act2}
 S(\varphi,\chi)  =  \int
\frac{d^2w}{2\pi} \left( \partial \varphi \bar \partial \varphi +
\partial \chi \bar
\partial \chi +
  \frac{\sqrt g}{4} {\cal R}(Q_\varphi \varphi + Q_{\tilde \chi}
  \tilde \chi) + \tilde \mu
     e^{\frac{2}{b}\varphi} - 2  e^{- \frac{1}{b}\, \varphi +
     i \sqrt k \, \tilde \chi}
\right) \, . \end{equation} In our derivation, the new action
$S(\varphi,\chi)$ arises as a perturbation of Liouville theory by
the exponential interaction term \eqref{int}. For the
exponentiation of our degenerate field insertions into a term of
the action it was crucial that we had replaced the Liouville
interaction by its dual one in the previous step. In fact, only
after the replacement $b \rightarrow b^{-1}$ in Liouville field
theory, the two interaction terms of eq.\ \eqref{act2} have a
common regime in which they both become small. A functional 
$S(\varphi,\chi)$ with the interaction \eqref{int} and the original Liouville
exponential $\exp (2b\varphi)$ was considered in
\cite{Giribet:2007uh} as a possible dual of the cigar conformal
field theory. The {\em twisted Sine-Liouville theory} such an 
 $S(\varphi,\chi)$ was meant to describe, however, is not really well-defined. It
certainly does not provide a weakly coupled dual for the strongly
curved cigar background.

\subsubsection{Relation with Sine-Liouville theory}

There is not much left to be done. In fact, in eqs.\ \eqref{cigL1}
and \eqref{act2}, we have derived a duality between the cigar
conformal field theory and some new 2-dimensional field theory
that involves two exponential interaction terms. It relates $N$
point functions between the two models and maps the strong
coupling regime of one model to the weak coupling regime of the
other. The only remaining problem is that our background charges
and interaction terms do not look at all like those of
Sine-Liouville theory.

Part of this issue can be repaired immediately. To do so, we observe
that the square length $Q^2 = Q_\varphi^2 + Q_{\tilde \chi}^2$ of
our background charge is the same as for Sine-Liouville theory,
i.e.\ $Q^2 = b^2$. Hence, it is possible to perform a rotation in
field space from the fields $\varphi$ and $\tilde \chi$ to some
new fields $\phi$ and $\tilde X$ with background charges $Q_\phi =
b$ and $Q_{\tilde X} = 0$, respectively, i.e.
\begin{eqnarray} \label{rot1}
\phi & = & (k-1) \varphi - i \sqrt{k} b^{-1} \tilde \chi \ \ ,
\\[2mm] \label{rot2}
\tilde X & = &  - i \sqrt{k}{b^{-1}}\,  \varphi - (k-1) \tilde \chi
\ \ .
\end{eqnarray}
When expressed through our new fields, the two exponential
interaction terms become
\begin{eqnarray} \label{VL}
V_L & = & \exp (2b ^{-1} \varphi) \ = \ \exp ( 2b ^{-1} (k-1) \phi
- 2 i
\sqrt{k}b^{-2} \tilde X) \ \ , \\[2mm] \label{Vp}
V_- & = & \exp (-b^{-1}\varphi +  i \sqrt {k} \tilde \chi) \ = \
\exp (b^{-1} \phi - i\sqrt{k} \tilde X) \ \ .
\end{eqnarray}
Note that the exponential $V_-$ is one term of the tachyon
potential \eqref{SLV} in Sine-Liouville theory. Only $V_L$ still
looks very different from the second contribution $V_+$ to the
tachyon potential. But we shall see below that $V_L$ may be
identified with $V_+$ through a reflection with respect to the
exponent of the interaction term $V_-$.

Before we explain the identification of $V_L$ and $V_+$ we want to
approach the issue of reflections in a more general context.
Suppose we are given some theory $S$ with $n$ bosonic fields. We
denote their background charges by $\vec{Q} = (Q^i)$ where $i = 1,
\dots , n$. Let us also assume that the $n$ fields interact
through $p$ exponential terms. These involve a sets of vectors
$\vec{\beta}_\nu = (\beta_{\nu}^i)$ with $\nu$ running from
$\nu=1$ to $\nu = p$. As in our example (\ref{VL}), (\ref{Vp}), we
shall assume $\vec{\beta}_{\nu} (\vec{Q} - \vec{\beta}_{\nu}) = 1$
so that all interaction terms are marginal. With these notations
introduced, our theory looks as follows,
\begin{equation} \label{actgen}
 S\ = \ \frac{1}{2\pi} \int d^2 w \left( \sum_{i=1}^n
\partial X_i \bar \partial X_i + \sum_{i=1}^n \frac{\sqrt{g}}{4}
 {\cal R} (\vec{Q},\vec{X}) + \sum_{\nu=1}^p \,
 \mu_\nu\, e^{2 (\vec{\beta}_\nu,  \vec{X})}\right)\ .
\end{equation}
 Now we can pass to an equivalent theory by performing one of
the following reflections
\begin{equation}\label{reflect} w_\rho:
\vec{\beta}_\nu \ \longrightarrow \
   \vec{\beta}_\nu + \vec{\beta}_\rho
+ (1-2 (\vec{\beta}_\nu,\vec{\beta}_\rho))
\frac{\vec{\beta}_\rho}{(\vec{\beta}_\rho,\vec{\beta}_\rho)} \
\ .
\end{equation}
In other words, we can pick any pair of labels $\nu, \rho \in 1,
\dots, p$ and then replace the vectors $\beta_\sigma$ in our
theory by
$$  \vec{\beta}'_\nu \ = \ w_\rho \vec{\beta}_\nu \ \ \ , \ \ \
   \vec{\beta}'_\sigma \ = \ \vec{\beta}_\sigma  \ \ \ \mbox{ for }
\sigma \ \neq\ \nu\ \ .  $$ The reflection of the vector
$\vec{\beta}_\nu$ should be accompanied by an appropriate
adjustment of the corresponding bulk coupling $\mu_\nu$. We shall
denote the corresponding coupling by $\mu'_\nu$. All other bulk
couplings $\mu'_\sigma = \mu_\sigma$ with $\sigma \neq \nu$ remain
the same. For $\nu = \rho$ the reflection invariance of $S$
follows from the self-duality of the Liouville field $X_\nu$. When
$\nu \neq \rho$, the equivalence of the corresponding models is a
consequence of a simple field identification (see Appendix \ref{reflection} for
more details).

Let us now apply these general remarks to the case at hand. After
the rotation \eqref{rot1}, \eqref{rot2}, our model is of the general
form \eqref{actgen} with
$$ \vec{\beta}_1\ = \ ((k-1)/b, - i\sqrt{k}/ b^2) \ \ \ , \ \ \
   \vec{\beta}_2\ = \ (1/2b , - i \sqrt{k}/2) $$
and $\vec{Q} = (b,0)$. We claim that a single reflection of
$\beta_1$ with $w_2$ is necessary in order to obtain the missing
interaction term of the Sine-Liouville model. Indeed,
$$ w_2(\vec{\beta}_1) \ = \ (1/2b, i\sqrt{k}/2)\ \ . $$
Hence, after reflection, our interaction term $\tilde \mu V_L$
takes the form
\begin{eqnarray}
\tilde \mu\, V_L \ = \ \tilde \mu e^{2\frac{k-1}{b} \, \phi - 2i
\frac{\sqrt{k}}{b^2} \tilde X} & = & - 2 \pi ^2 \lambda^2 \,
e^{\frac{1}{b}\phi + i \sqrt{k} \tilde X}\ = \ - 2 \pi ^2 \lambda^2
V_+ \\[2mm] \label{lambda}
\mbox{ where } \ \ \ \ \ \lambda^2 & = & \frac{\tilde \mu}{2\pi^2}
\ \frac{1}{\gamma(2-k)} \ \ . \end{eqnarray} Here we used the reflection
properties of tachyon vertex operators in $c=-2$ Liouville theory
(see Appendix \ref{reflection}). The value $c=-2$, and the precise form of the
new cosmological constant, is determined by the background charge
$Q_- = - \frac{i}{\sqrt 2}$ of the field
$ - \frac{i}{\sqrt{2}} (b^{-1} \phi - i \sqrt k \tilde X )$
in the exponent of $V_-$.

In order to make the coefficients of $V_+$ and $V_-$ in our final
answer look more symmetrically, we shift the zero mode of $\tilde
X$ such that we end up with
$$ S(\phi,X) \ = \ \frac{1}{2\pi} \int d^2w \left( \partial
\phi \bar \partial \phi + \partial X \bar \partial X +
  \frac{\sqrt g}{4} {\cal R} Q_\phi \phi  +
   2 \pi \lambda e^{\frac{1}{b}\, \phi + i \sqrt{k}\tilde X}
  + 2 \pi \lambda  e^{\frac{1}{b}\, \phi - i \sqrt{k}\tilde X}
    \right) \ \ .
$$
This is indeed the action of the Sine-Liouville model.
The parameter $\lambda$ is determined through $b$ by the two
equations \eqref{lambda} and \eqref{mutilde}.

It remains to address the precise form of the vertex operators
that we should use when calculating correlation functions for the
cigar through Sine-Liouville theory. In equation \eqref{cigL1},
these took the form
\begin{equation}
N^j_{m,\bar m} \, V_{\alpha} (z)
 \, V^\chi_{m- \frac{k}{2},\bar m - \frac{k}{2}}(z) \ = \
 \frac{\Gamma(- j - m)}{\Gamma(1+j + \bar m )}
  e^{ 2b(j+1 + \frac{1}{2b^2})  \varphi
 + i \frac{2}{\sqrt{k}}\left((m - \frac{k}{2}) \chi_L -
  (\bar m - \frac{k}{2})\chi_R\right) } ~ .
\end{equation}
Now we rewrite the exponents of these vertex operators in terms
of the rotated fields $\phi_L, \phi_R$ and $X_L,X_R$. The step
requires to spit the equations (\ref{rot1}), (\ref{rot2}) into four
equations for the left and right components of the various
fields. The resulting exponents are rather complicated,
\begin{eqnarray} \label{vert}
N^j_{m,\bar m} \, V_{\alpha} (z)
 \, V^\chi_{m- \frac{k}{2},\bar m - \frac{k}{2}}(z) & = &
  \frac{\Gamma(- j - m)}{\Gamma(1+j + \bar m )}
 \ e^{2 \alpha_\phi^L \phi_L + 2 \alpha_X^L X_L +
  2 \alpha_\phi^R \phi_R + 2 \alpha_X^R X_R} \\[2mm]
\label{alpha}
{\rm where} \ \ \  \left(\begin{array}{c}
   \alpha_\phi^L \\[4mm] \alpha_X^L
 \end{array} \right) & = &
 \left(\begin{array}{c}
    b(k-1)(j+1+ \frac{1}{2b^2}) +
     \frac{1}{b}(m-\frac{k}{2}) \\[4mm]
     - i \sqrt{k}(j+1+\frac{1}{2b^2})
       - i \frac{1}{\sqrt{k}}
      (k-1)(m-\frac{k}{2}) \end{array} \right)\ \ .
\end{eqnarray}
The parameters $\alpha_\phi^R$ and $\alpha_X^R$ are given by
similar formulas but with an opposite sign in the expression
for $\alpha_X^R$ and $\bar m$ instead of $m$. Now we perform
the reflection $w^L_2$ obtained from $\beta^L_2 = (1/2b,
- i\sqrt{k}/2)$ on the vector $\vec\alpha^L$,
$$ w^L_2(\vec\alpha^L) \ = \
   \vec\alpha^L + \vec \beta^L_2 + (1-2(\vec\alpha^L,
    \vec\beta_2^L))\frac{\vec{\beta^L_2}}{(\vec\beta^L_2,
\vec\beta^L_2)} \ = \ \left(b(j+1),im/\sqrt{k}\ \right)\ .
$$
The corresponding calculation for the right components differs
only by some signs and results in $w_2^R(\vec\alpha^R) =
(b(j+1),-i\bar m/\sqrt{k})$. It is remarkable that the reflection
$w_2$ maps the complicated expression \eqref{alpha} for the vector
$\vec \alpha$ onto something so much simpler. In particular, the
reflection removed the shifts $j \rightarrow j +1/2b^2$ and
$m \rightarrow m-k/2$ that entered our computations long ago
through the redefinitions \eqref{phired} and \eqref{Xred}.

The field identification of vertex operators also involves an
additional factor. This so-called reflection amplitude is worked
out in Appendix \ref{reflection}. In our case, it turns out to
remove the numerical prefactor in the vertex operator
\eqref{vert}, up to an overall sign. Namely, we find
\begin{eqnarray*}
N^j_{m,\bar m} \, V_{\alpha} (z)
 \, V^\chi_{m- \frac{k}{2},\bar m - \frac{k}{2}}(z)
& \sim & - \  e^{ 2 b (j + 1) \phi + i \frac{2}{\sqrt{k}}
 ( m X_L - \bar m X_R )} \\[2mm] & = &
   - \  e^{ 2 b (j + 1) \phi} e^{  i \frac{n}{\sqrt{k}}  X
 + i \sqrt{k} w \widetilde X
  }  \ \ ,
\end{eqnarray*}
where $m = (kw+n)/2$ and $\bar m = (kw-n)/2$, as before.
Hence, we recovered the
conventional vertex operators of Sine-Liouville theory.
Inserting our results into eq.\ \eqref{cigL1}, we obtain
\begin{align}
 \left \langle \prod_{\nu = 1}^N \Psi^{j_\nu}_{m_\nu , \bar m_\nu}
 \right \rangle^\cig\  = \  {\cal N}
  \left \langle \prod_{\nu = 1}^N
  e^{ 2 b (j_\nu + 1) \phi} e^{i\frac{n_\nu}{\sqrt{k}} X +
    i \sqrt{k} w_\nu  \widetilde X} \right\rangle^\SL
\end{align}
with the overall factor ${\cal N} =  ( - 1 )^{N-S} \pi^{N -2 - 2S}  \lambda^{- S}$ depending on $\lambda$ and $S$.
The right hand side of the above expression is to be evaluated
in Sine-Liouville theory with radius
$R = \sqrt{k}$ and a bulk
cosmological constant that is determined through $b$ by the
two equations \eqref{lambda} and \eqref{mutilde}. Thereby,
we have established the equality \eqref{correq} of
correlators in the two models on the sphere.

\section{Generalization to surfaces of higher genus}

Having successfully completed our proof of the FZZ-duality we
would now like to extend it to surfaces of genus $g \geq 1$. Most
of our analysis in subsection 3.2 carries over to general closed
Riemann surfaces without any changes. Our main task is to extend
the relation \eqref{cigL} between the cigar and Liouville field
theory. In order to do so, we will briefly review our previous
study \cite{HS} of $H^+_3$ correlation functions on higher genus
surfaces. At the same time, we shall include spectral flow. As in
the case of the sphere, we then descend to the cigar and derive a
relation with Liouville field theory. Some necessary background
material on how to gauge the $H^+_3$ WZNW model on higher genus
surfaces is collected in Appendix \ref{cigarg}. The final step in
the derivation of the correspondence between the cigar and
Liouville theory requires good control of the Jacobian for the
coordinate transformation from momenta $\mu_\nu$ etc. to insertion
points $y_i$. This Jacobian is discussed in the technical Appendix
\ref{jacobian}.

\subsection{The $H^+_3$ - Liouville correspondence - genus $g\geq 1$}
\label{Ribaultg}

{}From now on let $\Sigma$ be a generic Riemann surface of genus $g$
and with a fixed complex structure. On $\Sigma$ there exist $g$
holomorphic one-forms $\omega_l$ with $l=1,\cdots,g$. As usual, we
normalize them such that
\begin{align}
 \oint _{\alpha_k} \omega_l = \delta_{kl} ~,
 \qquad \oint_{\beta_k} \omega_l = \tau_{kl} ~,
\end{align}
where the set of $(\alpha_l,\beta_l)$ is a canonical basis of
homology cycles.  The complex matrix $\tau$ is the period matrix
of the surface $\Sigma$.

Let us turn attention to the fields $\beta, \gamma, \phi$ of the
WZNW model. Originally, these are (possibly multi-valued) functions
on the surface $\Sigma$. But we shall consider them as
(quasi-) periodic functions on the Jacobian by means of the Abel
map $(w_k) = (\int^w \omega_k) \in \mathbb{C}^g$. The periodicity
conditions we impose look as follows
\begin{align}
 & \beta ( w_k + \tau_{kl} n^l + m_k | \tau )
 \  =\  e^{2 \pi i n^l \lambda_l} \beta (w_k | \tau ) ~,
  \nonumber \\[3mm]
 & \gamma ( w_k + \tau_{kl} n^l + m_k | \tau )
  \ = \ e^{ - 2 \pi i n^l \lambda_l} \gamma (w_k | \tau ) ~,
\label{bcg} \\[2mm]
 & \phi ( w_k + \tau_{kl} n^l + m_k | \tau )
  \ = \ \phi (w_k | \tau ) + \frac{2 \pi n^l {\rm Im} \lambda_l}{b}
\nonumber
\end{align}
for $n^l,m_k \in {\mathbb Z}$. The complex parameters $\lambda_l,
l=1,\dots,g,$ that determine the behavior of $\beta,\gamma$ and
$\phi$ under shifts along the $\beta$-cycles are called {\em
twists}. Because of these twists, $\gamma$ does not possess a zero
mode. On the other hand $\beta$ still has $g-1$ zero modes. These
are proportional to $\lambda$-twisted holomorphic differentials
$\omega_\sigma^\lambda$ \cite{Bernard,HS}.

As in the genus zero case, we compute $(N+1)$-point function in
the presence of an insertion $v^S(\xi)$ of the spectrally flowed
identity field at $z=\xi$,
$$ \blangle \ \prod_{\nu=1}^{N} \, V_{j_\nu}(\mu_\nu|z_\nu)
  \, v^S (\xi) \
\brangle^H_{(\lambda,\varpi,\tau)} \ = \ \int{\cal{D}^\lambda}
\phi {\cal{D}^\lambda} \c {\tilde{\cal{D}}^\lambda} \b \
e^{-S[\phi,\c,\b]}\,  \prod_{\nu=1}^{N} \,
   V_{j_\nu}(\mu_\nu |z_\nu) \ e^{S\phi(\xi)/b} $$
on a Riemann surface $\Sigma$ of genus $g$. The subscript
$(\lambda,\varpi,\tau)$ indicates that we evaluate the correlator
with fixed twists $\lambda_k$, fixed coefficients $\varpi_\sigma$
of the $\beta$ zero modes, and fixed complex structure $\tau_{kl}$
on the Riemann surface. The evaluation of physical correlators in
the WZNW model requires setting $\lambda_k = 0$ and integrating
over zero modes $\varpi_\sigma$. But the construction of the
correlators in the gauged model (cigar) and other applications on
the WZNW model require to keep the explicit dependence on both
twists and zero modes (see below).

The calculation leading from the $H_3^+$ WZNW model to Liouville
field theory proceeds essentially as on the sphere before. It
utilizes a number of rather basic functions on the Jacobian that
we shall introduce while sketching the main steps of the
derivation.
See \cite{Fay,Mumford,AMV} for some properties of functions on a generic Riemann surface.
To begin with, we integrate out the field $\gamma(w)$,
just as in section 2. Due to the presence of the various vertex
operators $V_{j_\nu}(\mu_\nu|z_\nu)$, the field $\beta (w)$ takes
the following form
\begin{equation} \label{bspbg}
\beta (w) \ = \   \sum_{\nu=1}^N \mu_\nu \sigma_{\lambda} (w ,
z_\nu) + \sum_{\sigma=1}^{g-1} \varpi_\sigma \omega^\lambda_\sigma
(w) \ \ .
\end{equation}
This expression for $\beta(w)$ replaces our formula \eqref{beta}.
It involves the object $\sigma_{\lambda} (w , z_\nu)$ with
a single pole at $w = z_\nu$. The latter
may be constructed explicitly in terms of the theta function
\begin{align} \label{theta}
 \theta_\delta (z | \tau )
 \  = \ \sum_{n \in {\mathbb{Z}}^g}
  \, \exp i \pi [ (n + \delta_1 )^k \tau_{kl} (n + \delta_1 )^l
   + 2 (n+\delta_1 )^k ( z + \delta_2 )_k ]\  ~.
\end{align}
Here, $\delta_k = (\delta_{1k} , \delta_{2k} )$ with $\delta_{1k},
\delta_{2k} = 0, 1/2$ denotes the spin structure along the
homology cycles $\alpha_k$ and $\beta_k$. With the theta function
$\theta_\delta (z | \tau )$ we can build the following auxiliary
function $h_\delta(z)$ through
\begin{equation} \label{h}  ( h_\delta (z)
)^2 \ =\ \sum_k \, \partial_k
 \theta_{\delta} (0 |\tau) \omega_k^\lambda (z)\ \ .
\end{equation}
In terms of these objects, the propagator $\sigma_{\lambda} (w ,
z)$ can be written as \cite{HS}
\begin{equation} \sigma_{\lambda} (w , z) \ =\
\frac{(h_{\delta}(w))^2}{\theta_{\delta} (\int^w_z \omega)}
\frac{\theta_\delta (\lambda - \int^w_z \omega)}
     {\theta_\delta (\lambda )} \
     \label{tp}
\end{equation}
with an odd spin structure $\delta$. Thereby, we have fully
explained the general form \eqref{bspbg} of $\beta(w)$. Next, let us
see how to generalize the constraints \eqref{constraints} from the
sphere to an arbitrary surface. Because we inserted the operator
$v^S(\xi)$ in our correlator, the object $\beta(w)$ along
with its first $S-1$ derivatives has to vanish at $w = \xi$,
\begin{align} \label{constraintsg}
f_{n,\xi} (\mu,\varpi,\lambda) \ := \ \sum_{\nu=1}^N \mu_\nu
\sigma^{(n-1)}_{\lambda} (\xi , z_\nu) + \sum_{\sigma=1}^{g-1}
\varpi_\sigma \omega^{\lambda (n-1) }_\sigma (\xi) = 0 \ \ .
\end{align}
Here, the superscript $(n-1)$ stands for the $(n-1)^{th}$
derivative and the integer $n$ runs over $n=1,\dots,S$. In
contrast to the corresponding relations \eqref{constraints}, there
is no constraint for $n=0$, at least as long as the twists are
kept at generic values (see \cite{HS} for a more detailed
discussion).

Once more, we would like to bring the function \eqref{bspbg} into
a product form similar to eq.\ \eqref{bspb}. This may be achieved
using another basic fact about one-differentials on a surface of
genus $g$, namely that they possess $2(g-1)$ more zeros than they
possess poles. Consequently, we can rewrite $\beta (w)$ as
\begin{align}
 \beta (w) \ = \ u \frac{E(w,\xi)^S \prod_{i=1}^{M} E(w ,y_i)
 \sigma (w)^2} {\prod_{\nu=1}^N E(w,z_\nu)} ~. \label{condg}
\end{align}
This expression encodes the $M = N+2g-2-S$ zeroes of order one at
the points $w =y_i, i=1,\dots,M,$ and the zero of order $S$ at $w=
\xi$. It uses the well known prime form $E(z,w)$ which is
defined through
\begin{align} \label{prime}
 E (z , w )\  = \  \frac{\theta_\delta ( \int^z_w \omega | \tau)}
                   {h_\delta (z) h_\delta (w)} ~ ,
\end{align}
where $\theta_\delta$ and $h_\delta$ are the same as in eqs.\ \eqref{theta} and
\eqref{h} above. By construction, the prime from $E(z,w)$ has a single
zero at $z = w$. The other function $\sigma(w)$ that appears in
the formula \eqref{condg} is a $g/2$-differential with neither
poles nor zeros. Its definition can be found e.g.\ in
\cite{HS,VV}.

The rest of the calculation can be copied from our discussion in
section 2. As before, we redefine the field $\phi$ through the
following prescription,
\begin{align}
 \varphi (w,\bar w) & :=\
\phi (w,\bar w) + \frac{1}{2b} \left( S \ln |E(w , \xi) |^2
 + \right. \\
   & \qquad + \left.
   \sum_{i=1}^M \ln | E(w,y_i)|^2 \right.
    \left. - \sum_{\nu=1}^N \ln | E(w , z_\nu )|^2 + 2
    \ln | \sigma (w) |^2 - \ln |u \rho (w) |^2 \right) ~.
    \nonumber
\end{align}
Then we evaluate the change of the kinetic term. The resulting
formula for correlation functions in the WZNW model involves
Liouville correlators with $M = N+2g-2-S$ degenerate field
insertions. The precise expression is
\begin{align} \label{HLg}
& \blangle \ \prod_{\nu=1}^{N} \, V_{j_\nu}(\mu_\nu|z_\nu) \,  v^S
(\xi) \ \brangle^H_{(\lambda, \varpi,\tau)}  = \\[2mm] \nonumber
 & \qquad =\
\prod_{n=1}^S \delta ^2 (f_{ n,\xi} (\mu,\varpi,\lambda) )
|\sigma (\xi)|^{2S} |\Theta^g_N|^2
\nonumber \blangle \, \prod_{\nu=1}^N \, V_{\alpha_\nu} (z_\nu) \
\prod_{i=1}^{M} \, V_{-\frac{1}{2b}}(y_i)\, \brangle^{L}_\tau ~.
\end{align}
Our notations for fields in the Liouville correlation function on
the right hand side are the same as in eq.\ \eqref{R1} above. The
constraint functions $f_{n,\xi}$ were introduced in eq.\
\eqref{constraintsg}. In addition, the right hand side of eq.\
\eqref{HLg} involves a prefactor $\Theta^g_N$ of the form
\begin{align} \label{Thetag}
 |\Theta^g_N|^2
  &= e^{\frac{3}{4}kU_g} |\det {}' \nabla_\lambda|^{-2}
 |u |^{2 - 2 g - \frac{S}{b^2}}
  \prod_{\nu =1}^N |\sigma (z_\nu)|
  ^{ - 2 - \frac{2}{b^2}}
   \prod_{i = 1}^{M} |\sigma (y_i)|^{2 + \frac{2}{b^2}}\,  \times
   \\[2mm]
  &  \qquad \times \,
 \prod_{\mu < \nu}^N | E ( z_\mu , z_\nu ) |^{\frac{1}{ b^2}}
      \prod_{i < j}^{M} | E (y_i , y_j )|^{\frac{1}{b^2}}
      \prod_{\nu=1}^N \prod_{k=1}^{M}
       |E (z_\nu , y_k)|^{- \frac{1}{b^2}}\  ~. \nonumber
\end{align}
The prime in  $\det {}' \nabla_\lambda$ indicates that we drop the
contribution from the zero mode. The function $U_g$, finally, is
defined by
\begin{align}
 U_g \ = \  \frac{1}{192\pi^2} \int d^2 w d^2 y
 \sqrt{g(w)} {\cal R} (w)\sqrt{g(y)} {\cal R} (y)
 \ln | E(w,y)|^2 ~.
\end{align}
Here, $g(w)$ denotes the metric on the Riemann surface and ${\cal
R}$ is its curvature. This concludes our derivation of the
$H^+_3$-Liouville correspondence for higher genus surfaces. The
special case $S=0$ was treated in more detail in \cite{HS}.

\subsection{The cigar-Liouville correspondence - genus $g \geq 1$}
\label{reduction}

In this subsection, we would like to relate correlators of cigar
model to those of Liouville field theory with a free boson.
Thereby, we shall extend eq.\ \eqref{cigL} to a general Riemann
surface of genus $g$. Our starting point is the following
presentation of the cigar correlation functions in terms of
correlators of the $H^+_3$ WZNW model and a free boson $X$,
\begin{align} \label{startg}
  \left\langle \prod_{\nu = 1}^N
 \Psi^{j_\nu}_{m_\nu ,\bar m_\nu } (z_\nu ) \right\rangle
 ^\text{cig} =  \Delta_{\rm FP} (\hat {\cal A})   \int {\cal D} g
   {\cal D} X \prod_{l=1}^{g} d^2 \lambda_l
    e^{- S^{\text{cig}}[g,X]_\lambda }
 \prod_{\nu = 1}^N
 \Psi^{j_\nu}_{m_\nu ,\bar m_\nu } (z_\nu ) ~.
\end{align}
The vertex operators are given in eq.~\eqref{parafield},
and formula
\eqref{startg} is derived in Appendix \ref{cigarg}. The right hand
side is computed in the product of an $H_3^+$ WZNW model and a
free bosonic field theory,
\begin{align}
  &\left\langle \prod_{\nu = 1}^N
 \Psi^{j_\nu}_{m_\nu ,\bar m_\nu } (z_\nu ) \right\rangle
 ^\text{cig} \ = \  |\det {}' \partial|^2 \int \prod_{\sigma=1}^{g-1} d ^2
\varpi_\sigma
  \prod_{l=1}^g d^2 \lambda_l
   \, \times \\[2mm]  &\qquad \times \nonumber
 \left[ \ \prod_{\nu = 1 }^N
 \int \frac{d^2 \mu_\nu }{|\mu_\nu |^2}
  \mu_\nu^{m_\nu} \bar {\mu}_\nu^{\bar m_\nu} \right]  \left\langle
   V^X_{-\frac{kS}{2},-\frac{kS}{2}}  (\xi)\,  e^{S \phi(\xi)/b}
 \prod_{\nu = 1}^N V^X_{m_\nu,\bar m_\nu}(z_\nu)
  V_{j_\nu} (\mu_\nu | z_\nu )
\right \rangle^{H\times F}_{S}  ~.
\end{align}
Since the vertex operator does not include $(b,c)$-ghosts,
the Faddeev-Popov determinant can be factored out.
Here we have chosen the measure for $\varpi_\sigma$
such that the overall factor becomes simple.

Utilizing the result \eqref{HLg} from the previous subsection, we
can express all $H_3^+$ correlators through correlation functions
in the Liouville field theory. As in the case $g=0$, we redefine
the field $X$ to remove the remaining explicit $\mu$-dependence,
\begin{align}
  \chi_L(w,\bar w)  & := X_L(w,\bar w)  \label{xg}
   - i \frac{\sqrt{k}}{2} \left( S \ln E(w,\xi) +
\right. \\[2mm] &  \hspace*{1cm}
\left.  + \sum_{i=1}^{M} \ln E(w,y_i)- \sum_{\nu = 1}^N \ln E(w,z_\nu) + 2 \ln \sigma (w) - \ln u
\rho (w) \right)
  ~.  \nonumber
\end{align}
A similar redefinition is performed for $\chi_R$. {}From the
definition of $X$ (see eq.\ \eqref{gaugex} in Appendix
\ref{cigarg}) we can see that $X_L$ and $X_R$ receive shifts
similar to the one for the  field $\phi$ when we go around a
$\beta$-cycle, see the third line of eq.\ \eqref{bcg}. Through
the redefinition
\eqref{xg}, the new fields $\chi_L$ and $\chi_R$ become periodic.
The mechanism is the same as for the Liouville field $\varphi$. We
can now proceed as before and obtain
\begin{align}
  \left \langle \prod_{\nu = 1}^N \Psi^{j_\nu}_{m_\nu , \bar
m_\nu} (z_\nu )
 \right  \rangle^\text{cig}
  & = \
\int \frac{\prod_{j=1}^{M} d ^2 y_j}{M!}
  \prod_{\nu = 1}^N  N^{j_\nu}_{m_\nu , \bar m_\nu}
    \times \nonumber \\[2mm] \label{cLg}
& \times \,   \left \langle\  \prod_{\nu=1}^N V_{\alpha_\nu} (z_\nu  )
 \, V^\chi_{m_\nu- \frac{k}{2},\bar m_\nu - \frac{k}{2}}(z_\nu)\
  \prod_{j =1}^{M} V_{- \frac{1}{2b}} (y_j  )
 \, V^\chi_{\frac{k}{2}, \frac{k}{2}}(y_j)\, \right  \rangle ~.
\end{align}
The derivation of eq.\ \eqref{cLg} requires a generalization of
the expression \eqref{jacobxi} for the Jacobian to surfaces of
genus $g \geq 1$,
\begin{align}
\label{jacob}
& \prod_{\nu=1}^N \frac{d^2 \mu_\nu}{|\mu_\nu|^2}
 \prod_{\sigma=1}^{g-1} d^2 \varpi_\sigma  \prod_{l=1}^g d ^2 \lambda_l
 \prod_{n=1}^S \delta(f_{n,\xi} (\mu ,\varpi, \lambda ))
  \frac{| \det {}' \partial |^{2}}{| \det {}' \nabla_\lambda |^{2}} = \\[2mm] \nonumber
 & \, = \,
  \frac {\prod_{\mu < \nu}^{N} | E(z_\mu , z_\nu ) |^2
    \prod_{i<j}^{M}| E(y_i, y_j) |^2
 \prod_{i=1}^{M} |\sigma(y_i)|^2 }
{\prod_{\nu=1}^N \prod_{i=1}^{M} | E(z_\nu , y_i) |^2
   \prod_{\nu=1}^{N} | \sigma(z_\nu) |^2 | \sigma(\xi) |^{2S}}
   \frac{d^2 u}{|u|^{4 -2g + 2S}}
 \prod_{i=1}^{M} d^2 y_i
 ~.
\end{align}
We prove this formula in Appendix \ref{jacobian}. Once we have
arrived at eq.\ \eqref{cLg}, the steps we performed in section 3.2
go through without any changes. In particular, we can exponentiate
the degenerate field insertions and then work our way through
reflections and rotations until we arrive at the relation
\eqref{correq} between correlators on arbitrary surfaces.

\section{Conclusion and open problems}
\def\del{\partial}

In the previous three sections we have established complete
agreement between correlation functions of tachyon vertex
operators on the cigar and in Sine-Liouville theory. On
the other hand, equivalence of the two models, i.e.\ the
agreement of {\em all} correlation functions, still needs
to be addressed. The proof is only completed once we have
shown that both models possess the same chiral symmetry
and that our tachyon vertex operators form the set of
primary fields with respect to this chiral algebra. Both
statements are in fact well established. Therefore, we shall
only outline the main ingredients and provide a few
references to the original literature.

The chiral symmetry of the cigar conformal field theory, often
denoted by $\widehat{\cal W}_\infty(k)$, was studied many years
ago, right after the model had been first discussed. A very
convincing description of $\widehat{\cal W}_\infty(k)$ along
with extensive references to earlier contributions can be
found in \cite{Bakas:1991fs}. Given the basic fields $\phi$
and $X$ of the cigar conformal field theory one may construct
the following parafermionic currents
\begin{equation} \label{paraf}
\Psi_\pm(z) \ = \
i \left( b^{-1} \, \del \phi \pm i \sqrt{k}\, \del X \right)\
  e^{\pm 2i \frac{1}{\sqrt{k}} \, X_R} \ \ .
\end{equation}
By the equation of motion, these fields turn out to be
chiral. Since their construction involves splitting the field
$X$ into its chiral components, however, parafermionic currents
are not local. On the other hand, $\Psi_\pm(z)$ may be used
to generate an infinite set $W_s, s=2,3,4,\dots$ of local
chiral fields with weight $h_s =s$. Through repeated operator
products one first finds the usual stress energy tensor $T =
W_2$ and then a field $W_3$ of the form
\begin{equation} \label{W3}
 W_3(z) \ = \ \frac{6k-8}{3}\, (\del X)^3 +
                \frac{2}{b^2}\, (\del \phi)^2 \del X +
                \frac{1}{b^3}\, \del^2\phi\, \del X -
                \frac{k}{b}\, \del\phi\,  \del^2 X +
                \frac{k}{6} \del^3 X \ \ .
\end{equation}
One may show that all higher weight fields $W_s, s \geq 4$, can be
recovered from operator products of $T$ and $W_3$ alone.
We can therefore think of $\widehat{\cal W}_\infty(k)$ either
as the chiral algebra obtained from $T$ and $W_3$ or as the
algebra of local fields with parafermionic constituents. The
algebra $\widehat{\cal W}_\infty(k)$ suffices to generate the
entire state space of the cigar conformal field theory out of
the primaries $|j,m,\bar m\rangle \ = \ \Psi^J_{m,\bar m}(0)
|0\rangle$.

Let us now switch to Sine-Liouville theory. We may employ
either the explicit formula \eqref{paraf} for parafermions
or the construction \eqref{W3} of $W_3$ to show that the
chiral symmetry algebra $\widehat{\cal W}_\infty(k)$ is
preserved by the interaction terms in the Sine-Liouville
model, i.e.\
\begin{equation}
\oint_z dw\, W_3(w) V_\pm(z,\bar z) \ = \ 0 \ = \ \oint_z dw\,
 \Psi_\epsilon (w)  V_\pm(z,\bar z)\ .
\end{equation}
Here, $\epsilon = \pm$ and $V_\pm$ are the two exponentials
that appear in the interaction terms of the Sine-Liouville
model. Interested readers may find a more detailed
discussion and references to earlier works in \cite{Fateev:2005kx}.
In conclusion, the cigar and Sine-Liouville models possess
the same chiral symmetries. Therefore, the main result of this
note proves that they are equivalent.

In the introduction we have presented the AdS/CFT correspondence
as our main motivation for studying the strong-weak coupling
duality of the non-compact cigar geometry. Obviously, the
2-dimensional cigar is a rather simple toy model for realistic,
higher dimensional holographic backgrounds, such as $AdS_5 \times
S^5$. Still, it is intriguing to see how the physics of a strongly
curved holographic background can be mapped to a dual weakly
coupled world-sheet model. Let us stress that none of the steps in
our analysis seem to rely in an essential way on the particular
target space dimension of the cigar. On the other hand, we
certainly exploited the extended chiral symmetry of the model. It
seems unlikely, however, that chiral symmetries are really all
that crucial. In \cite{Fateev:1995ht,Fateev:1996ea}, for example,
Fateev described several dualities similar to the one between the
cigar and Sine-Liouville theory, but involving massive integrable
models. In any case, finding explicit higher dimensional examples
of weakly coupled world-sheet models for strongly curved
holographic backgrounds appears as an interesting direction for
further research. Such dual models could eventually mediate
between strongly coupled string physics and a weakly coupled gauge
theory on the boundary of $AdS_5$.

Even though the extension of our analysis to higher dimensional
target spaces seems possible in principle, it could be technically
challenging. The space $AdS_5 \times S^5$, for example, arises as
a base of the coset superspace SU(2,2$|$4)/(SO(4,1)$\times$SO(5)).
Since (super-)groups of higher rank are involved in this
construction of an $AdS_5$ background, the AdS/CFT correspondence
motivates an extension of the correspondence between $H^+_3$ and
Liouville theory to {(super-)}groups such as SL(N) or PSL(N$|$N).
WZNW models on SL(N), for example, possess a well studied relation
with Toda field theories through Hamiltonian reduction. Encouraged
by the successful treatment of SL(2), one may hope to upgrade
embeddings of Toda theory into WZNW models to a full
correspondence. At the critical level $k=N$, such a relation is
understood as one of the ramifications of Langlands duality. 
Interested readers may find a detailed explanation of the so-called
{\em geometric} Langlands duality and its relation to conformal
field theory, along with many further references e.g.\ in
\cite{Frenkel:2005pa} (see also \cite{RT,Giribet:2008ix} for the
connection with the $H^+_3$-Liouville correspondence). We hope to
report on an off-critical version of the geometric Langlands
duality for SL(N) and other (super-)groups of higher rank in the
future.
\bigskip
\bigskip
\bigskip

\noindent {\bf Acknowledgements:} We would like to thank Patrick
Dorey, Vladimir Fateev, Amit Giveon, Sylvain Ribault, Peter R\o
nne, Samson Shatashvilli and J\"org Teschner for useful
discussions, remarks and comments on the manuscript. The work of
YH is supported in part by Research Fellowships of the Japan
Society for the Promotion of Science for Young Scientists.

\appendix
\section{Cigar field theory as a gauged WZNW model}
\label{cigarg}

In this appendix we discuss the construction of the cigar
conformal field theory as a gauged WZNW model, see also, e.g,
\cite{GK,Gawedzki,Martinec:1991ea,Dijkgraaf:1991ba}. Our treatment
is not restricted to the case of genus $g=0$. To begin with, let us
recall that the action of the $H^+_3$ WZNW model takes the form
\begin{align}
 S^{\text{WZNW}} [g]\ =\ \frac{k}{2\pi} \int d^2 w \left(
  \partial \phi \bar \partial \phi + e^{-2 \phi}
  \bar \partial \gamma  \partial \bar \gamma \right)\ .
\end{align}
Here and in the following we shall use the letter $g$ as a
shorthand for the fields $\gamma,\bar \gamma$ and $\phi$. Upon
introduction of the two auxiliary fields $\beta, \bar \beta$ we
may recover the first order action~\eqref{action} we have used
throughout the main text, except for a different normalization of
the field $\phi$.

The action of the coset theory is obtained through the usual
prescription. If we decompose the U(1) valued gauge field $A$
through ${\cal A} = A dw + \bar A d \bar w$, the action of the
gauged model becomes
\begin{align}
S^{\text{cig} } [g,{\cal A}]\  = \ \frac{k}{2 \pi} \int d^2 w
\left[
  (\bar \partial \phi + \bar A) (\partial \phi + A )
   + e^{- 2 \phi} (\bar \partial + \bar A) \gamma
     (\partial + A) \bar \gamma \right] ~.
\end{align}
Expectation values of any operator ${\cal O}$, such as a product
of tachyon vertex operators, for example, are now computed through
the associated path integral
\begin{align}
 \langle {\cal O} \rangle ^{\text{cig}} \ = \ \frac{1}{V_{\rm sym}}
 \int {\cal D} g {\cal D} {\cal A} \  e^{- S^{\text{cig}} [g,{\cal A}] }
 \ {\cal O}
 ~, \label{cigarpi}
\end{align}
where $V_{\rm sym}$ is the volume of the gauge group. Path
integrals of this form may be evaluated with the help of the
Faddeev-Popov prescription, by introducing an auxiliary system of
$(b,c)$-ghosts. To this end insert
\begin{align}
 1 \ = \ \Delta_{\rm FP} ({\cal A}) \int [d x d \alpha \prod_{k=1}^g
 d ^{2} \lambda _k ]\
  \delta ({\cal A} - \hat {\cal A}^{\alpha } (x,\lambda) )
  \label{fp}
\end{align}
into our path integral \eqref{cigarpi}. Here, $\Delta^{\rm
FP}(\cal A)$ is the Faddeev-Popov determinant and the gauge field
is parametrized as
\begin{align}
 \hat {\cal A}^{\alpha} (x ,\lambda )
 \ =\  d x + * d \alpha
 - \pi i \bar \lambda_k (\tau_2^{-1})^{kl} \omega_l
 - \pi i \lambda_k (\tau_2^{-1})^{kl} \bar \omega_l ~,
\end{align}
where $k,l =1,\cdots,g$ and $(\tau_2)_{kl} = {\rm Im}
\,\tau_{kl}$. On a sphere we can always choose $A = \partial (x +
i \alpha)$ and $\bar A = \bar \partial (x - i \alpha)$. This
choice is  locally possible on a generic Riemann surface of genus
$g$, but globally we have to include zero modes. These
zero modes correspond to the possibly non-vanishing holonomies
along the various cycles, and they are proportional to
holomorphic one-form $\omega_l$
with $l=1,\cdots g$ on the Riemann surface. As our notation
suggests, the parameter $\lambda_k$ turns out to be the twist
along non-contractible cycles. In \cite{HS}, we fixed these twists
to obtain the relation between correlators of $H_3^+$ model and
Liouville theory. In our present context, however, we have to
integrate out $\lambda_k$ as well. Insertion of the identity
\eqref{fp} leads to
\begin{align}
 \langle {\cal O} \rangle ^{\text{cig}} \ = \ \frac{1}{V_{\rm gauge}}
 \int [d x d \alpha \prod_{k=1}^g
 d ^{2} \lambda_k d g ] \, \Delta_{\rm FP}\,  (\hat {\cal A}^{\alpha })
\,  e^{- S^{\text{cig}} [g,\hat {\cal A}^{\alpha }] } \  {\cal O}
 \label{fpcorr}
\end{align}
after the integration over ${\cal A}$. Since we can show that
$\Delta_{\rm FP} (\hat {\cal A}^{\alpha })$ and $S^{\text{cig}}
[g,\hat {\cal A}^{\alpha }]$ are independent of $\alpha $, the
integration over $\alpha $ only gives an overall factor $V_{\rm
sym}$, as long as  the inserted vertex operators are independent
of $\alpha$ as well.

Our first aim is to evaluate the Faddeev-Popov measure. Since
the variation of gauge field is given by
\begin{align}
 \delta \hat {\cal A} \ =\  d \delta x + *d \delta \alpha  -
 \pi i \delta \bar \lambda_k(\tau_2^{-1})^{kl} \omega_l
 - \pi i \delta \lambda_k (\tau_2^{-1})^{kl} \bar \omega_l ~,
\end{align}
the inverse of the measure becomes
\begin{align}
 \Delta^{-1}_{\rm FP} (\hat {\cal A})
 &= \  \int [d \delta x d \delta \alpha
 \prod_{k=1}^g d ^{2} \delta \lambda _k ]
  \delta (\delta \hat {\cal A} )  \\[2mm]
 &= \  \int [d ^2 \beta ' d \delta x d \delta \alpha
 \prod_{k=1}^g d ^{2} \delta \lambda_k ]
  \exp \left[ 2 \pi i \int d^2 w \left(
\beta ' \bar \partial (\delta x + i \delta\alpha)
  + \bar \beta ' \partial (\delta x - i \delta \alpha)   \right) \right]
 \nonumber \\[2mm] & \qquad \qquad\qquad \qquad  \times
   \exp \left[  2 \pi i \int d^2 w \left(
 \pi \bar \beta_0 \delta \bar \lambda_k (\tau_2^{-1})^{kl} \omega_l
 +\pi  \beta_0 \delta \lambda_k (\tau_2^{-1})^{kl} \bar \omega_l
  \right) \right]~. \nonumber
\end{align}
We invert this expression following the standard trick. Thereby,
we can express the Faddeev-Popov measure through an path integral
over fermionic ghost systems $(b,c)$ and $(\bar b,\bar c)$ along
with a Grassmann integral over $2g$ additional variables $\xi_k$
and $\bar \xi_k$. The latter are associated with the variations
$\delta \lambda_k$ and $\delta \bar \lambda_k$. Thus, the measure
takes the form
\begin{align}
 \Delta_{\rm FP} (\hat {\cal A})
 &=  \  \int [d ^2 b d ^2 c
 d ^{2g} \xi ]
  \ e^{ \left[ - \frac{1}{2\pi} \int d^2 z \left(
b \bar \partial c
  + \bar b \partial \bar c  \right) +  \int \left(
 \bar b_0 \bar \xi_k ( \tau_2^{-1})^{kl} \omega_l
 + b_0 \xi_k (\tau_2^{-1})^{kl} \bar \omega_l
  \right) \right]} \nonumber \\[2mm]
 & =  \int [d ^2 b d ^2 c ] e^{ - S_{\rm gh} [b,c]}
  \prod_{k=1}^{g} \left | \int
  \bar b_0 ( \tau_2^{-1})^{kl} \omega_l \right |^2 ~.
  \label{fpfinal}
\end{align}
In passing to the second line, we have performed the integration
over $\xi_k$ and $\bar \xi_k$. Correlation functions are now
obtained from \eqref{fpcorr} by inserting the expression of
Faddeev-Popov measure \eqref{fpfinal}. In particular, if the
operator ${\cal O}$ does not involve any ghosts, we can explicitly
perform the integration over the $(b,c)$-ghost system, which leads to
$$ \int [d ^2 b d ^2 c ] e^{ - S_{\rm gh} [b,c]} \ = \
 |{\det}' \partial|^2\  \det \tau_2^{-1}\ \ . $$
One may think of the determinant $|\det '\partial|^2$ as arising
from the Jacobian $|d {\cal A} / d x d \alpha|$. The factor $\det
\tau_2^{-1}$, on the other hand, is due to our normalization of
parameter $\lambda_k$.

It now remains to evaluate $S^{\text{cig}} [g,\hat {\cal
A}^{\alpha = 0} (x,\lambda) ]$. Following \cite{GK}, we may
re-express this action through a $H^+_3$ WZNW model and an
additional free boson. Let us separate $x$ as $x = x_L + x_R$ with
the condition $x_L = (x_R)^*$. Then the gauge field $A$ can be
written as
\begin{align}
 A  \ = \ \partial x_L  -
 \pi i \lambda_k ( \tau_2 ^{-1})^{kl} \omega_l
 \ = \ ( \Gamma_\lambda e^{x_L})^{-1}
  \partial ( \Gamma_\lambda e^{x_L}  )
\end{align}
with
\begin{align}
\Gamma_\lambda (w , \bar w)\  =\
 e^{ - \pi i \lambda_k (\tau_2 ^{-1})^{kl}
 ( \int^w \omega_l (z) - \int^{\bar w} \bar \omega_l (\bar z) )} ~.
\end{align}
If the arguments are translated along the various non-trivial
cycles of our surface, the factor $\Gamma_\lambda$ picks up the
following phases
\begin{align}
\Gamma_\lambda (w_k + \tau_{kl} n^l + m_k | \tau)\ = \
 e^{ - 2 \pi i n^l \lambda_l} \Gamma_\lambda (w_k|\tau)\ \ .
\end{align}
Even though the anti-holomorphic part of $\Gamma_\lambda$ does not
contribute to the chiral gauge field $A$, it is required for
$\Gamma_\lambda$ to possess good shift properties. A similar
representation can be written down for the component $\bar A$ of
the gauge field. With this in mind we now redefine our fields
according to
\begin{align}
 \phi + \frac{1}{2}
 (x_L + x_R + \ln |\Gamma_\lambda|^2) \mapsto \phi   ~, \qquad
\bar \Gamma_\lambda e^{x_R} \gamma \mapsto \gamma ~, \qquad
\Gamma_\lambda e^{x_L} \bar \gamma \mapsto \bar \gamma ~.
\end{align}
In terms of the new field, the action becomes a sum of two simple
contributions,
\begin{align}
 S^{\text{cig}} [g,{\cal A}^{\alpha = 0} (x,\lambda ) ]
 \ = \ S^{\text{WZNW}} [g]_\lambda +
 \frac{1}{2\pi} \int d^2 w \bar \partial X \partial X ~.
 \label{sep}
\end{align}
The index $\lambda$ on the WZNW action indicates that the WZNW
model is defined with the twist as in \eqref{bcg}. Furthermore, we
defined the free boson $X = X_L + X_R$ by
\begin{align}
X_L \ :=\  - \frac{\sqrt{k}}{2} i (x_L + \ln \Gamma_\lambda) ~,
\qquad X_R \ := \  \frac{\sqrt{k}}{2} i (x_R + \ln \bar
\Gamma_\lambda)
 ~ . \label{gaugex}
\end{align}
The chiral components of $X$ also satisfy non-trivial boundary
conditions, due to the shift with $\ln \Gamma_\lambda$. In
summary, we have shown that the action for the cigar model is
given by summing the action of a $\lambda$-twisted $H_3^+$ WZNW
model, a free boson $X$, and a $(b,c)$-system. In the main text,
we only consider situations in which our insertion ${\cal O}$ does
not involve fields $b$ and $c$. Therefore, the $(b,c)$-system
decouples from the rest of the theory.

\section{Reflection: Self-duality and field identification}
\label{reflection}

This appendix collects a few fact about Liouville theory, its
self-duality and reflection property. These are applied to
multi-field models of the form \eqref{actgen} in the second
subsection.

\subsection{Field identification in Liouville field theory}

Let us first consider a single Liouville field $\omega$ with bulk
cosmological constant $\rho$ and background charge $Q_\omega = d +
1/d$. Our aim is to describe the reflection coefficient of fields
in this theory, i.e.\ the function $D(\alpha^L,\alpha^R)$ that
features in the relation
\begin{equation}\label{reflD}  V_{\alpha^L,\alpha^R} \
= \ D(\alpha^L,\alpha^R) V_{Q_\omega-\alpha^L,Q_\omega-\alpha^R}
\end{equation}
between vertex operators $V_{\alpha^L ,\alpha^R} = \exp (2
\alpha^L \omega _L + 2 \alpha^R \omega _R)$. Here, we shall allow
for situations in which the exponent contains contributions from
the dual Liouville field $\tilde \omega$, i.e.\ with $\alpha_L
\neq \alpha_R$. The reflection coefficient $D$ is severely
constrained by the operator product of generic vertex operators
with degenerate ones as
\begin{align}
 &V_{-\frac{d}{2},-\frac{d}{2}} (z)\,  V_{\alpha^L,\alpha^R} (0)
 \\[2mm]
 &  \qquad \sim
 z^{d\alpha^L} \bar z^{d\alpha^R}\,
V_{\alpha^L - \frac{d}{2},\alpha^L - \frac{d}{2} } (0)
 + C_- (\alpha^L,\alpha^R)
 \, z^{d(Q_\omega-\alpha^L)} \bar z^{d(Q_\omega-\alpha^R)}\,
 V_{\alpha^L + \frac{d}{2},\alpha^R + \frac{d}{2}}(0) + \dots
 \  .\nonumber
\end{align}
Combining this expansion with the reflection equation
\eqref{reflD}, we assign the following two conditions
\begin{align} \begin{split}
  & C_- (\alpha^L,\alpha^R)
 D(\alpha^L + \tfrac{d}{2},\alpha^R + \tfrac{d}{2})
 \ = \ D(\alpha^L,\alpha^R) \ , \\[2mm]
 & D(\alpha^L,\alpha^R)
 D(Q_\omega - \alpha^R , Q_\omega - \alpha^L)  \ =\  1 \ .
\label{const}
\end{split}
\end{align}
The coefficient $C_-$ may be determined through a free field
computation, as e.g.\ in \cite{Dotsenko},
\begin{align}
 C_- (\alpha^L,\alpha^R) &=\  - \frac{\rho}{2 \pi} \int d^2 x
  \langle V_{\alpha^L,\alpha^R} (0)
  V_{-\frac{d}{2},-\frac{d}{2}}(1)
 e^{2d\omega(x)}
 V_{Q_\omega-\alpha^L -\frac{d}{2},Q_\omega-\alpha^R -\frac{d}{2}}(\infty)
  \rangle  \nonumber \\[2mm]
  &= \ - \rho  \gamma(1+d^2)
  \frac{\Gamma(-1+2d\alpha^L-d^2)\Gamma(1-2d\alpha^R)}
 {\Gamma(2 - 2d\alpha^R+d^2)\Gamma(2d\alpha^L)}\  .
\end{align}
There is a unique analytic solution to the constraints
\eqref{const} that is consistent with the duality symmetry under
simultaneous exchange $d \leftrightarrow 1/d$ and $\alpha^L
\leftrightarrow \alpha^R$. It is given by
\begin{align} \label{D}
 D(\alpha^L , \alpha^R ) \ = \ ( \rho \gamma(d^2))^{(Q_\omega-\alpha^L - \alpha^R)/d}
 \frac{\Gamma(2d\alpha^L - d^2)
 \Gamma (-1+ \frac{2\alpha^R}{d}-\frac{1}{d^2})}
 {d^2 \Gamma(1- 2d\alpha^R + d^2)
 \Gamma (2 - \frac{2\alpha^L}{d} + \frac{1}{d^2})}\  .
\end{align}
Applying the reflection to the Liouville field itself, i.e.\ to
the case with $\alpha^L = \alpha^R = d$, we infer that the
bulk cosmological constant $\tilde \rho$ of the dual Liouville
field theory must take the form
\begin{align} \label{reflbcc}
 \tilde \rho  \  = \ \gamma ^{-1} (1/d^2)
  (\rho \gamma(d^2))^{1/d^2}\  .
\end{align}
In the main text, the reflection of vertex operators is performed
in a Liouville field theory with parameter $d= i/\sqrt2$ and with
bulk cosmological constant $\rho = - 2$. If we insert these values
into our general formula for the reflection amplitude, we obtain
\begin{align}
D(\alpha^L , \alpha ^R ) \ =\  - \frac{\Gamma(1-i\sqrt2 \alpha^R)}{\Gamma(i\sqrt2
\alpha^L)}\  \ \ .
\end{align}
Formulas \eqref{D} and \eqref{reflbcc} contain all the information
that is needed to perform reflections of the type \eqref{reflect}
we considered in section 3.2.

\subsection{Reflection in theories with several bosonic fields}
\def\va{{\vec \alpha}}
\def\vb{{\vec \beta}}

Our notations and conventions in this subsection are the same as
in subsection 3.2.3 of the main text. Let us suppose that $\vec
\beta$ is one of the vectors satisfying $(\vec \beta,\vec Q - \vec
\beta) =1$. We want to analyze the field identification for a
vertex operator $V_{\vec \alpha}$ that is induced by the Liouville
interaction $\exp 2(\vec\beta,\vec X)$. Here $\vec \alpha$ can be
any vector. It is convenient to redefine
\begin{align}
 X_{\vec \beta} =  (d^{-1}_{\vec \beta} \vec \beta , \vec X) ~, \qquad
 d_{\vec \beta} = \sqrt{(\vec \beta , \vec \beta)} ~.
 \label{dbeta}
\end{align}
The background charge for this bosonic field $X_{\vec \beta}$
is $q_{\vec \beta} = d_{\vec \beta}  + d_{\vec \beta}^{-1}$
and the interaction term is now
$\exp 2 d_{\vec \beta} X_{\vec \beta}$.

To begin with, let us isolate from the vector $\vec
\alpha$ its component along $\vec \beta$,
\begin{equation}
\va \ = \ \alpha_{\vec \beta} ( d^{-1}_{\vec \beta} \vec \beta )
  + \left( \va -
\frac{(\va,\vb)}{(\vb,\vb)} \vb\right)\ , \qquad
\alpha_\vb \ :=\ d^{-1}_{\vec \beta} (\vec
\alpha,\vb)  \ .
\end{equation}
The reflection along $\vb$ is controlled by the value of the
background charge $q_\vb$ along $\vb$. Hence, upon reflection, the
vector $\va$ gets replaced by
\begin{equation} \label{reflproof}
w_\vb(\va)\, =\,  (q_{\vec \beta} - \alpha_{\vec \beta} )
 ( d^{-1}_{\vec \beta} \vec \beta ) + \left( \va -
\frac{(\va,\vb)}{(\vb,\vb)} \vb\right) \, = \, \va + \vb +
(1-2(\va,\vb)) \frac{\vb}{(\vb,\vb)}\ .
\end{equation}
Formula \eqref{reflproof} is used twice in the main text, namely
in eq.\ \eqref{reflect} and after eq.\ \eqref{alpha}.

The results of the previous subsection may also be used to
determine the reflection amplitude that is needed to relate
$V_\va$ with its reflection $V_{w(\va)}$.
We then find
\begin{equation}
 D(\vec \alpha ^L , \vec \alpha ^R ) \ = \ ( \mu_\vb \gamma(d_\vb^2))^{(q_\vb-\alpha^L_\vb - \alpha^R_\vb)/d_\vb}
 \ \frac{\Gamma(2d_\vb\alpha^L_\vb - d^2_\vb)\,
 \Gamma (-1+ \frac{2\alpha^R_\vb}{d_\vb}-\frac{1}{d^2_\vb})}
 {d^2_\vb \, \Gamma(1- 2d_\vb\alpha^R_\vb + d^2_\vb)
 \, \Gamma (2 - \frac{2\alpha^L_\vb}{d_\vb} + \frac{1}{d^2_\vb})}\  .
\end{equation}
Here, $\mu_\vb$ is the bulk cosmological constant in front of the
interaction term $\exp 2(\vb,\vec X)$. In passing from an
interaction term $\exp 2(\vb,\vec X)$ to the dual one, we must
replace
\begin{equation}
\vb \ \longrightarrow \ \frac{\vb}{(\vb,\vb)} \ \ \ \ \ \text{and}
\ \ \ \ \mu_\vb \ \longrightarrow \ \tilde \mu_\vb \ = \ \tilde
\gamma^{-1}(1/d^2_\vb) \ (\mu_\vb \gamma(d^2_\vb))^{1/d^2_\vb}\  .
\end{equation}
The expression for the dual cosmological constant was obtained
from eq.\ \eqref{reflbcc} by inserting the value $d_\vb$ defined
in eq.\ \eqref{dbeta}.

\section{On the Jacobian}
\label{jacobian}

The aim of this appendix is to compute the Jacobian \eqref{jacob}
that arises when we change variables from the momenta $\mu_\nu$ to
insertion points $y_i$. We will first explain the main ideas in
the case of the sphere. Thereby, we rederive eq.\ \eqref{jacobxi}
that was already established in \cite{RT,Ribault}. Our derivation
generalizes more or less directly to surfaces of higher genus $g
\geq 1$, and these will be treated in the second subsection.

\subsection{The Jacobian on the sphere}

For the sphere with genus $g=0$, the separation of variables
\eqref{bspb} may be written in terms of the individual momenta
$\mu_\nu$ by comparing residues,
\begin{align}
 \mu_\nu \ = \ u \frac{(z_\nu - \xi)^S \prod_{i=1}^{N-2-S} (z_\nu - y_i)}
 {\prod_{\mu \neq \nu =1}^N (z_\nu - z_\mu)}\ .
 \label{ssv}
\end{align}
Thereby, we obtain the following
relation between differentials
\begin{align}
 \frac{d \mu_\nu}{\mu_\nu} \ =\
 \frac{d u}{u} - \sum_{i=1}^{N-2-S} \frac{d y_i}{(z_\nu - y_i )} ~.
\end{align}
Before we continue, let us set $S=0$. We shall treat the more
general case with $S\neq 0$ a bit later. The measure in momentum
space may read
\begin{align}
 \prod_{\nu = 1}^{N}  \frac{d^2 \mu_\nu}{|\mu_\nu|^2 }
 \delta^2 (\sum_{\nu} \mu_\nu)
 \ = \ \prod_{\nu = 1}^{N-1}
    \left|  \frac{d u}{u} - \sum_{i=1}^{N-2} \frac{d y_i}
 {(z_\nu - y_i )}
    \right|^2
   \left[
\frac{\prod_{\mu=1 }^{N-1}  |z_N - z_{\mu} |^2}
   {|u|^2 \prod_{i=1}^{N-2} |z_N - y_i|^2} \right] ~ .
  \label{jacob_mid}
\end{align}
We would like to rewrite the first factor on the right hand side.
In order to do so, we observe that it may be expressed through the
correlation function of an auxiliary $(b,c)$-system. If we adjust
the central charge such that the $b$ and $c$ possess conformal
weight $h_b = 1$ and $h_c = 0$ and furthermore normalize the
fields according to $c(z) b(y) \sim 1/(z - y)$, we find
\begin{align}
\prod_{\nu = 1}^{N-1}
    \left(  \frac{d u}{u} - \sum_{i=1}^{N-2} \frac{d y_i}{(z_\nu - y_i  )}
    \right)
    \ = \ \left\langle \prod_{\nu = 1}^{N-1} c (z_\nu) \prod_{i=1}^{N-2} b (y_i)
      \right\rangle \frac{du}{u} \prod_{i=1}^M (-dy_i) ~.
\end{align}
Notice that one of the insertions $c(z_\nu)$ is replaced by the
zero mode, i.e., a constant mode. Utilizing the usual bosonization
formulas for $(b,c)$-systems we obtain
\begin{align}
\left| \left\langle \prod_{\nu = 1}^{N-1} c (z_\nu)
\prod_{i=1}^{N-2} b (y_i)
      \right\rangle \right |^2
\ = \ \frac{\prod_{\mu < \nu = 1}^{N-1} |z_\mu - z_\nu|^2
  \prod_{i <j = 1}^{N-2} |y_i - y_j|^2}
 {\prod_{\nu=1}^{N-1} \prod_{i=1}^{N-2} |z_\nu - y_i|^2} ~.
\end{align}
When this result is inserted back into eq.~\eqref{jacob_mid}, we
recover a special case of the Jacobian \eqref{jacobxi} with $S=0$,
\begin{align}
 \prod_{\nu = 1}^{N}  \frac{d^2 \mu_\nu}{|\mu_\nu|^2}
 \delta^2 (\sum_{\nu} \mu_\nu)
 \ = \ \frac{\prod_{\mu < \nu = 1}^{N} |z_\mu - z_\nu|^2
  \prod_{i <j = 1}^{N-2} |y_i - y_j|^2}
 {\prod_{\nu=1}^{N} \prod_{i=1}^{N-2} |z_\nu - y_i|^2}
 \frac{d ^2 u}{|u|^{4}}
 \prod_{i=1}^{N-2} d^2 y_i ~.
 \label{jacob0}
\end{align}

In order to treat the remaining cases with $S \neq 0$, we perform
an induction in $S$. So, let us assume that the relation
\eqref{jacobxi} holds for $S=s$. If $S$ is increased to $S =s+1$,
the left hand side of eq.\ \eqref{jacobxi} reads
\begin{align}
\text{lhs\eqref{jacobxi}} \ = \
  \prod_{\nu=1}^N \frac{d^2 \mu_\nu}{|\mu_\nu|^2}
  \delta ^2 \left( \beta_{-s - 1} (\mu_\nu ) \right)
 \prod_{n=0}^s \delta ^2 \left(\sum_\nu \frac{\mu_\nu}
 { (\xi - z_\nu )^n} \right)\ ,
\end{align}
where $\beta_{-s-1} = \sum_\nu \mu_\nu (\xi - z_\nu)^{- s -1}$
contains the contributions of the $N$ source terms to the mode
$\beta_{-s-1}$ of $\beta$. The right hand side of eq.\
\eqref{jacobxi} can be obtained from the case $S=s$ by choosing
one of the insertion points and moving it to the position $\xi$.
Without loss of generality, we shall take $x := y_{N-2-s} \to
\xi$. This gives
\begin{align} \nonumber
\lim_{x\rightarrow \xi}(\text{rhs\eqref{jacobxi}}_{S=s}) & = \
\frac{d^2 u}{|u|^{4 + 2s}}
 \prod_{i=1}^{N-3-s} d^2 y_i d^2 x \ \delta^2
 \left( \beta_{-s - 1} (y_i,x ) \right)  \times \\[2mm] \nonumber
 &  \hspace*{2mm} \times  \frac{\prod_{\mu < \nu}^{N}
   |z_{\mu} - z_{\nu}|^2
    \prod_{i<j}^{N-3-s} |y_{i} - y_{j}|^2  \prod_{i=1}^{N-3-s} |y_{i} -x|^2}
  {\prod_{\nu=1}^N \prod_{i=1}^{N-3-s} |z_\nu - y_i|^2
\prod_{\nu=1}^N |z_\nu - x|^2}
 ~.
\end{align}
Thereby, we have reduced our problem to showing that
\begin{align}
 \delta ^2 \left( \beta_{-s - 1} (y_j,x ) \right)
  =  \frac{1}{|u|^2}\frac{\prod_{\nu=1}^N |z_\nu - \xi|^2}
 {\prod_{i=1}^{N-3-s} |y_{i} -\xi|^2} \delta ^2 (x - \xi)\ .
 \label{d2d}
\end{align}
For $S=s$, the function $\beta(w)$ is known to take the form
\begin{align}
 \beta (w) = u \frac{(w-\xi)^s (w-x) \prod_{i=1}^{N-3-s}(w-y_i)}
                    {\prod_{\nu=1}^N (w - z_\nu) } ~.
\end{align}
We can take this expression and expand around $x \sim \xi$ to
obtain the following expression for the mode $\beta_{-s-1}$
\begin{align}
\beta_{-s-1} \ =\ u \frac{ \prod_{i=1}^{N-3-s}(\xi-y_i)}
                    {\prod_{\nu=1}^N (\xi - z_\nu) }
                    (\xi - x)
\end{align}
in terms of $y_i$. This equation leads to eq.\ \eqref{d2d}, and
thereby establishes that the formula eq.~\eqref{jacobxi} for the
Jacobian is valid for all $0 \leq S \leq N-2$.

\subsection{The Jacobian for genus $g\geq 1$}

We now repeat the steps of the previous subsection in the case of
generic genus $g \geq 1$. In this case, the separation of
variables takes the form
\begin{align}\label{SOVg}
 \sum_{\nu=1}^N \mu_\nu \sigma_{\lambda} (w ,
z_\nu) + \sum_{\sigma=1}^{g-1} \varpi_\sigma \omega^\lambda_\sigma
(w) \ = \ u \frac{ E(w,\xi) ^ S
 \prod_{i=1}^{M} E(w ,y_i) \sigma (w)^2}
 {\prod_{\nu=1}^N E(w,z_\nu)} ~.
\end{align}
Here we have denoted $M=N-2g-2-S$, as before.
{}From this equation we can deduce the formula for the
momenta $\mu_\nu$ as
\begin{align*}
 \mu_\nu \ = \ u \frac{E(z_\nu,\xi)^S
 \prod_{i=1}^{M} E(z_\nu ,y_i) \sigma (z_\nu)^2}
 {\prod_{\mu \neq \nu}^N E(z_\nu,z_\mu)} ~, \qquad
 \frac{d \mu_\nu}{\mu_\nu} \ = \ \frac{du}{u} + \sum_{i = 1}^M
 \partial_{y_i} \ln E ( z_\nu , y_i)  d y_i ~.
\end{align*}
This expression strongly indicates that the Jacobian could
be derived with the use of $(b,c)$-ghost system since
the propagator of a $(b,c)$-system on a surface of genus $g$
can be expressed through the prime form as $\langle c(z) b(y) \rangle = \partial_y \ln E(z,y)$.

In the case of genus zero
only the integral over the momenta $\mu_\nu$ appears
in the left hand side of \eqref{jacobxi},
but in the case of generic genus the integrals over the twists $\lambda_l$
and over the zero modes $\varpi_\sigma$ are involved in
\eqref{jacob} as well.
For the twists $\lambda_l$ we utilize the relations to $y_i$ as
\begin{align}
 \lambda_l \ = \ S \int_w^{\xi}  \omega_l  +
  \sum_{i=1}^M \int_w^{y_i} \omega_l
- \sum_{\nu = 1}^N  \int_w^{z_\nu} \omega_l - 2
\int^{\Delta}_{(g-1)w} \omega_l  ~, \label{lambdac}
\end{align}
where $\Delta$ denotes the Riemann constant.
By acting with the differential $d$ on these equations, we obtain
the following simple relations as
\begin{align}
 d \lambda_l = \sum_{i=1}^M \omega_l (y_i) d y_i ~.
 \label{dlambda}
\end{align}
For the zero modes $\varpi_\sigma$ we use the general expression as
\begin{align} \label{dzero}
 &\sum_{\nu=1}^N d \mu_\nu \sigma_{\lambda} (\eta_\rho ,
z_\nu) + \sum_{\sigma=1}^{g-1} d\varpi_\sigma
\omega^\lambda_\sigma (\eta_\rho)
\\[2mm]
& \qquad \quad =\ u \frac{ E(\eta_\rho,\xi) ^ S
 \prod_{i=1}^{M} E(\eta_\rho ,y_i) \sigma (\eta_\rho)^2}
 {\prod_{\nu=1}^N E(\eta_\rho ,z_\nu)} \left[ \frac{du}{u}+
 \sum_i \partial_{y_i} E(\eta_\rho ,y_i) dy_i \right]~, \nonumber
\end{align}
which is deduced from eq.~\eqref{SOVg}.
Here we set $w=\eta_\rho$ as an arbitrary point on the
Riemann surface $\Sigma$.

As in the previous subsection we start from $S=0$ case and then
generalize to $S \neq 0$ case by making use of the induction
procedure. One may worry about the measure of zero modes
as the expression in the left hand side of
\eqref{dzero} involves $d \mu_\nu$ in addition to $d \varpi_\sigma$.
This problem can be resolved by the following simple observation,
\begin{align}
  \prod_{\nu=1}^N \frac{ d \mu_\nu }{ \mu_\nu }
 \prod_{\rho=1}^{g-1}
  & \left[\sum_{\nu=1}^N d \mu_\nu \sigma_{\lambda} (\eta_\rho ,
z_\nu)   + \sum_{\sigma=1}^{g-1} d\varpi_\sigma
\omega^\lambda_\sigma (\eta_\rho )\right]  \prod_{l=1}^g
d\lambda_l
 \ = \\[2mm] \nonumber & \qquad =\
 \prod_{\nu=1}^N \frac{ d \mu_\nu }{ \mu_\nu }
 \prod_{\sigma = 1}^{g-1}  d \varpi _\sigma
 \prod_{l=1}^g d\lambda_l
 \det _{\sigma, \rho}\omega_\sigma^\lambda (\eta_\rho ) ~.
\end{align}
Combining everything obtained above, the left hand side
of \eqref{jacob} can be rewritten as
\begin{align}
& \prod_{\nu=1}^N \frac{ d \mu_\nu }{ \mu_\nu }
 \prod_{\sigma=1}^{g-1} d \varpi_\sigma
 \prod_{l=1}^g d\lambda_l
  \ = \ \frac{1}{ \det _{\sigma, \rho}\omega_\sigma^\lambda (\eta_\rho )}
\prod_{\nu=1}^N \left[ \frac{du}{u} + \sum_{i = 1}^M
 \partial_{y_i} \ln E ( z_\nu , y_i)  d y_i \right]
   \\[2mm]  & \times
\prod_{\rho=1}^{g-1} \left\{ u \frac{
 \prod_{i=1}^{M} E(\eta_\rho ,y_i) \sigma (\eta_\rho)^2}
 {\prod_{\nu=1}^N E(\eta_\rho,z_\nu)}
\left[ \frac{du}{u} + \sum_{i = 1}^M
 \partial_{y_i} \ln E ( \eta_\rho , y_i)  d y_i \right] \right\}
\prod_{l=1}^g \left[ \sum_{i=1}^M \omega_l (y_i) d y_i \right] ~.
\nonumber
\end{align}
Once again it is
advantageous to express the right hand side of the previous
equality through correlators in an auxiliary $(b,c)$-system.
Notice that $b(y)$ has $g$ zero modes $\omega_l(y)$ for
genus $g$. Moreover, there is a single constant mode for $c(w)$ (see,
e.g., \cite{VV}). With these facts and the propagator of
$(b,c)$-system, we can express the measure as
\begin{align} \label{jacobbc}
 &\prod_{\nu=1}^N \frac{ d ^2 \mu_\nu }{ |\mu_\nu|^2 }
 \prod_{\sigma=1}^{g-1}  d ^2 \varpi_\sigma
 \prod_{l=1}^g d ^2 \lambda_l
 \ = \  \frac{1}
 { |\det _{\sigma, \rho}\omega_\sigma^\lambda (\eta_\rho )|^2}
 \frac{1}{| \det ' \partial |^2 }
\times  \\[2mm] & \qquad \times   \prod_{\rho =1}^{g-1}
   \left\{\frac{
 \prod_{i=1}^{M} |E(\eta_\rho ,y_i)|^2 |\sigma (\eta_\rho) |^4}
 {\prod_{\nu=1}^N |E(\eta_\rho,z_\nu)|^2} \right\}
\left|
 \left\langle   \prod_{\sigma = 1}^{g-1} c(\eta_\sigma)
 \prod_{\nu=1}^ N c(z_\nu)
 \prod_{k=1}^{M}
 b(y_k)  \right\rangle \right|^2
 \frac{d^2 u}{|u|^{4-2g}} \prod_{i=1}^{M} d^2 y_i \nonumber ~.
\end{align}
The factor $1/|\det '\partial|^2$ is included to
divide the contribution from the partition function.

In the following we will show that eq.~\eqref{jacobbc} is indeed
equal to eq.~\eqref{jacob} by
utilizing the bosonization formulas of $(b,c)$-systems.
First, we rewrite the correlation function of $(b,c)$-ghosts
in \eqref{jacobbc} as \cite{VV}
\begin{align}
& \left|
 \left\langle   \prod_{\sigma = 1}^{g-1} c(\eta_\sigma)
 \prod_{\nu=1}^ N c(z_\nu)
 \prod_{k=1}^{M}
 b(y_k)  \right\rangle \right|^2
 \ = \  \frac{1}{| \det {}' \partial  |}
| \theta ( \sum_\sigma \eta_\sigma + \sum_\nu z_\nu - \sum_i y_i  +
\Delta) |^2
   \times \label{C18}\\[2mm] & \times
  \frac { \prod_{\sigma < \rho}^{g-1}| E(\eta_\sigma , \eta_\rho ) |^2
  \prod_{\sigma =1}^{g-1} \prod_{\nu=1}^N | E(\eta_\sigma , z_\nu ) |^2
 \prod_{\mu < \nu}^{N} | E(z_\mu , z_\nu ) |^2
    \prod_{i<j}^{M}| E(y_i, y_j) |^2
 \prod_{i=1}^{M} |\sigma(y_i)|^2 }
{\prod_{\sigma=1}^{g-1} \prod_{i=1}^{M} | E(\eta_\sigma , y_i) |^2
\prod_{\nu=1}^N \prod_{i=1}^{M} | E(z_\nu , y_i) |^2
   \prod_{\rho=1}^{g-1} | \sigma (\eta_\rho) |^2  \prod_{\nu=1}^{N} | \sigma(z_\nu) |^2 }
 ~ . \nonumber
\end{align}
The factor $( \det {}' \partial  \bar \partial )^{-1/2}$ is
the partition function of a complex boson. The theta function, which
may be written as $|\theta (\sum_\sigma \eta_\sigma - \Delta - \lambda)|$
by means of eq.\ \eqref{lambdac}, arises from summing over the solitonic modes.
Notice that the factors involving $E(\eta_\sigma,z_\nu)$ and $E(\eta_\sigma,
y_i)$ are canceled if we insert the above expression \eqref{C18} into
eq.\ \eqref{jacobbc}. In this way, the entire dependence on $\eta_\sigma$
resides in a single factor that is independent on any of the variables.
Since the $\eta_\sigma$-dependence is expected to drop out in the end,
the cancellation of terms involving both $y_i,z_\nu$ and $\eta_\sigma$ is
an important intermediate step.

In order to incorporate the last factor on the left hand side of
our formula \eqref{jacob}, we need to analyze the partition
function with twists $\lambda_k$. Up to now we worked with a
$(b,c)$-ghost system without twists, but this does not yield any
partition function with $\lambda$-dependence. Therefore, we shall
now deal with $(b,c)$-ghosts with twists $\lambda_k$, where the
ghosts satisfy the same twisted boundary conditions as the
$(\beta,\gamma)$-system. Consequently, the $b$-ghost has $g-1$
zero modes which are proportional to the $g-1$ twisted holomorphic
one-differentials $\omega^\lambda_\sigma (w)$. The simplest
non-zero correlator in the twisted $(b,c)$-system is
\begin{align} \label{partb}
 \left\langle \prod_{\rho =1}^{g-1} b(\eta_\rho)
 \right\rangle_\lambda
  \ = \  \det {}' \nabla_\lambda   \det _{\sigma,\rho}
 \omega^\lambda_\sigma (\eta_\rho) \ .
\end{align}
We show now that this function is useful to remove the
$\eta_\sigma$-dependence in eq.~\eqref{jacobbc}.
Again application of the usual bosonization formulas leads to
\begin{align} \label{C22}
 \left|\left\langle \prod_{\rho =1}^{g-1} b(\eta_\rho)
 \right\rangle\right|^2
 \ = \ \frac{1}{|\det {}' \partial|}
   | \theta ( \sum_\sigma \eta_\sigma - \Delta - \lambda) |^2
  \prod_{\sigma < \rho = 1}^{g-1} | E(\eta_\sigma , \eta_\rho) |^2
  \prod_{\sigma = 1}^{g-1} | \sigma (\eta_\sigma) |^2 ~.
\end{align}
Then the combination with eq.\ \eqref{partb} gives the equality
\begin{align*}
 |\det {}' \nabla_\lambda |^2 \ = \
  \frac{1}{|\det _{\sigma,\rho}
 \omega^\lambda_\sigma (\eta_\rho)|^2}  \frac{1}{|\det {}' \partial|}
   | \theta ( \sum_\sigma \eta_\sigma - \Delta - \lambda) |^2
 \prod_{\sigma < \rho = 1}^{g-1} | E(\eta_\sigma , \eta_\rho) |^2
\prod_{\sigma = 1}^{g-1} | \sigma (\eta_\sigma) |^2  \ .
\end{align*}
This equality removes the all
$\eta_\sigma$-dependent terms and at the same time leads to
eq.~\eqref{jacob} for $S=0$.

The cases with $S \neq 0$ are treated as in the previous
subsection, i.e.\ by induction in $S$. Therefore, we assume that
the Jacobian is of the anticipated form when $S=s$ and try to
establish the same for $S = s+1$. The first few steps are
performed in precisely the same way as on the sphere. They lead to
the following requirement
\begin{align}\label{d2dg}
 \delta^2 (\beta_{-s-1}(y_i,x))
 \ =\ \frac{1}{|u|^2} \frac{\prod_{\nu=1}^N |E(\xi , z_\nu )|^2}
  {\prod_{i=1}^{N-2g-3-s} |E (\xi , y_i ,)|^2|\sigma(\xi)|^4}
  \delta ^2 (x - \xi)
\end{align}
that replaces our formula \eqref{d2d} from the previous
subsection. We may prove this equation by recalling that the prime
form behaves as $E(\xi, x) \sim \xi - x$ for $\xi \sim x$.
Therefore, the mode expansion of $\beta(w)$ around $w \sim \xi$
gives
\begin{align}
 \beta_{-s-1} \ = \ u \frac{\prod_{i=1}^{N-2g-2-s}
 E(\xi , y_i)\sigma(\xi)^2}
  {\prod_{\nu=1}^{N}E(\xi , z_\nu)} (\xi - x) ~.
\end{align}
With this result we can easily deduce first eq.\ \eqref{d2dg} and
then the anticipated expression \eqref{jacob} for the Jacobian
with $S=s+1$ from the case $S=s$. Thereby, we conclude our
derivation of the Jacobian \eqref{jacob}.


\begin{thebibliography}{99}
\bibitem{Giveon:1994fu}
  A.~Giveon, M.~Porrati and E.~Rabinovici,
  ``Target space duality in string theory,''
  Phys.\ Rept.\  {\bf 244}, 77 (1994)
  [arXiv:hep-th/9401139].

\bibitem{Elitzur:1991cb}
  S.~Elitzur, A.~Forge and E.~Rabinovici,
  ``Some global aspects of string compactifications,''
  Nucl.\ Phys.\  B {\bf 359}, 581 (1991).

\bibitem{Mandal:1991tz}
  G.~Mandal, A.~M.~Sengupta and S.~R.~Wadia,
  ``Classical solutions of two-dimensional string theory,''
  Mod.\ Phys.\ Lett.\  A {\bf 6}, 1685 (1991).

\bibitem{Witten:1991yr}
  E.~Witten,
  ``On string theory and black holes,''
  Phys.\ Rev.\  D {\bf 44}, 314 (1991).

\bibitem{KKK}
  V.~Kazakov, I.~K.~Kostov and D.~Kutasov,
  ``A matrix model for the two-dimensional black hole,''
  Nucl.\ Phys.\  B {\bf 622}, 141 (2002)
  [arXiv:hep-th/0101011].

\bibitem{FZZ}
 V.A.~Fateev, A.B.~Zamolodchikov and Al.B.~Zamolodchikov,
 unpublished.

\bibitem{Buscher1}
  T.~H.~Buscher,
  ``A symmetry of the string background field equations,''
  Phys.\ Lett.\  B {\bf 194}, 59 (1987).

\bibitem{Buscher2}
  T.~H.~Buscher,
  ``Path integral derivation of quantum duality in nonlinear sigma models,''
  Phys.\ Lett.\  B {\bf 201}, 466 (1988).


\bibitem{FH}
  T.~Fukuda and K.~Hosomichi,
  ``Three-point functions in Sine-Liouville theory,''
  JHEP {\bf 0109}, 003 (2001)
  [arXiv:hep-th/0105217].


\bibitem{Hori:2001ax}
  K.~Hori and A.~Kapustin,
  ``Duality of the fermionic 2d black hole and ${\cal N} = 2$ Liouville theory as
  mirror symmetry,''
  JHEP {\bf 0108},  045 (2001)
  [arXiv:hep-th/0104202].

\bibitem{RT}
  S.~Ribault and J.~Teschner,
  ``$H_3^+$ WZNW correlators from Liouville theory,''
  JHEP {\bf 0506}, 014 (2005)
  [arXiv:hep-th/0502048].

\bibitem{HS}
  Y.~Hikida and V.~Schomerus,
  ``$H^+_3$ WZNW model from Liouville field theory,''
  JHEP {\bf 0710}, 064 (2007)
  [arXiv:0706.1030 [hep-th]].

\bibitem{Fateev}
  V.~Fateev, unpublished.

\bibitem{Maldacena:2000hw}
  J.~M.~Maldacena and H.~Ooguri,
  ``Strings in $AdS_3$ and SL(2,{$\mathbb R$}) WZW model. I: The spectrum,''
  J.\ Math.\ Phys.\  {\bf 42}, 2929  (2001)
  [arXiv:hep-th/0001053].

\bibitem{Ribault}
  S.~Ribault,
  ``Knizhnik--Zamolodchikov equations and spectral flow in $AdS_3$ string
  theory,''
  JHEP {\bf 0509}, 045 (2005)
  [arXiv:hep-th/0507114].

\bibitem{MO3}
  J.~M.~Maldacena and H.~Ooguri,
  ``Strings in $AdS_3$ and the SL(2,${\mathbb R}$) WZW model. III: Correlation  functions,''
  Phys.\ Rev.\  D {\bf 65}, 106006 (2002)
  [arXiv:hep-th/0111180].

\bibitem{Zamolodchikov:1995aa}
  A.~B.~Zamolodchikov and A.~B.~Zamolodchikov,
  ``Structure constants and conformal bootstrap in Liouville field theory,''
  Nucl.\ Phys.\  B {\bf 477}, 577 (1996)
  [arXiv:hep-th/9506136].

\bibitem{Schomerus:2005aq}
  V.~Schomerus,
  ``Non-compact string backgrounds and non-rational CFT,''
  Phys.\ Rept.\  {\bf 431}, 39 (2006)
  [arXiv:hep-th/0509155].

\bibitem{Ponsot:1999uf}
  B.~Ponsot and J.~Teschner,
  ``Liouville bootstrap via harmonic analysis on a noncompact quantum  group,''
  arXiv:hep-th/9911110.

\bibitem{Teschner:2003en}
  J.~Teschner,
  ``A lecture on the Liouville vertex operators,''
  Int.\ J.\ Mod.\ Phys.\  A {\bf 19S2}, 436 (2004)
  [arXiv:hep-th/0303150].

\bibitem{Teschner:2001rv}
  J.~Teschner,
  ``Liouville theory revisited,''
  Class.\ Quant.\ Grav.\  {\bf 18}, R153 (2001)
  [arXiv:hep-th/0104158].

\bibitem{Giribet}
  G.~Giribet,
  ``The string theory on $AdS_3$ as a marginal deformation of a linear  dilaton
  background,''
  Nucl.\ Phys.\  B {\bf 737}, 209 (2006)
  [arXiv:hep-th/0511252].

\bibitem{Giribet:2007uh}
  G.~Giribet and M.~Leoni,
  ``A twisted FZZ-like dual for the 2D black hole,''
  arXiv:0706.0036 [hep-th].

\bibitem{Bernard}
  D.~Bernard,
  ``On the Wess-Zumino-Witten models on Riemann surfaces,''
  Nucl.\ Phys.\  B {\bf 309}, 145 (1988).


\bibitem{Fay}
  J.~Fay,
  ``Theta functions on Riemann surfaces,''
  Lecture Notes in Mathematics 352,
  Springer-Verlag (1973).

\bibitem{Mumford}
  D.~Mumford,
  ``Tata lectures on theta, Vols. I, II,''
  Progress in Mathematics 43,
  Birkh\"auser (1984)

\bibitem{AMV}
  L.~Alvarez-Gaume, G.~W.~Moore and C.~Vafa,
  ``Theta functions, modular invariance, and strings,''
  Commun.\ Math.\ Phys.\  {\bf 106}, 1 (1986).

\bibitem{VV}
  E.~P.~Verlinde and H.~L.~Verlinde,
  ``Chiral bosonization, determinants and the string partition function,''
  Nucl.\ Phys.\  B {\bf 288}, 357 (1987).

\bibitem{Bakas:1991fs}
  I.~Bakas and E.~Kiritsis,
  ``Beyond the large $N$ limit: Non-linear $W_\infty$ as symmetry of the SL(2,${\mathbb R}$)/U(1) coset model,''
  Int.\ J.\ Mod.\ Phys.\  A {\bf 7S1A}, 55 (1992)
  [Int.\ J.\ Mod.\ Phys.\  A {\bf 7}, 55 (1992)]
  [arXiv:hep-th/9109029].

\bibitem{Fateev:2005kx}
  V.~A.~Fateev and S.~L.~Lukyanov,
  ``Boundary RG flow associated with the AKNS soliton hierarchy,''
  J.\ Phys.\ A  {\bf 39}, 12889 (2006)
  [arXiv:hep-th/0510271].



\bibitem{Fateev:1995ht}
  V.~A.~Fateev,
  ``The duality between two-dimensional integrable field theories and sigma
  models,''
  Phys.\ Lett.\  B {\bf 357}, 395 (1995).

\bibitem{Fateev:1996ea}
  V.~A.~Fateev,
  ``The sigma model (dual) representation for a two-parameter family of
  integrable quantum field theories,''
  Nucl.\ Phys.\  B {\bf 473}, 509 (1996).


\bibitem{Frenkel:2005pa}
  E.~Frenkel,
  ``Lectures on the Langlands program and conformal field theory,''
  arXiv:hep-th/0512172.


\bibitem{Giribet:2008ix}
  G.~Giribet, Y.~Nakayama and L.~Nicolas,
  ``Langlands duality in Liouville-$H_3^+$ WZNW correspondence,''
  arXiv:0805.1254 [hep-th].

\bibitem{GK}
  K.~Gawedzki and A.~Kupiainen,
  ``Coset construction from functional integrals,''
  Nucl.\ Phys.\  B {\bf 320}, 625 (1989).

\bibitem{Gawedzki}
  K.~Gawedzki,
  ``Non-compact WZW conformal field theories,''
  arXiv:hep-th/9110076.

\bibitem{Martinec:1991ea}
  E.~J.~Martinec and S.~L.~Shatashvili,
  ``Black hole physics and Liouville theory,''
  Nucl.\ Phys.\  B {\bf 368} (1992) 338.

\bibitem{Dijkgraaf:1991ba}
  R.~Dijkgraaf, H.~L.~Verlinde and E.~P.~Verlinde,
  ``String propagation in a black hole geometry,''
  Nucl.\ Phys.\  B {\bf 371}, 269 (1992).

\bibitem{Dotsenko}
  V.~S.~Dotsenko,
  ``Lectures on conformal field theory,'' Advanced Studies in Pure Mathematics
 {\bf 16}, 123 (1988).



\end{thebibliography}
\end{document}